\documentclass[lettersize,journal]{IEEEtran}
\usepackage{amsmath,amsfonts}
\usepackage{array}
\usepackage[caption=false,font=normalsize,labelfont=sf,textfont=sf]{subfig}
\usepackage{textcomp}
\usepackage{booktabs}
\usepackage{stfloats}
\usepackage{url}
\usepackage{verbatim}
\usepackage{graphicx}
\usepackage{MnSymbol}
\usepackage{physics}
\usepackage{tikz,pgfplots}
\usepackage{lipsum,adjustbox}
\usepackage{mathtools}
\usetikzlibrary{external} 
\usepackage{caption}
\usepackage{algorithm, algpseudocode}
\usetikzlibrary{calc}
\usepackage{hyperref}
\usepackage{multirow} 
\usepackage{siunitx}
\usepackage{svg}

\newtheorem{remark*}{Remark}

\usetikzlibrary{arrows.meta}
\usepackage[style=ieee, maxbibnames=99]{biblatex}

\usepackage{pifont}% http://ctan.org/pkg/pifont
\newcommand{\cmark}{\ding{51}}%
\newcommand{\xmark}{\ding{55}}%
\begin{document}

\title{\textcolor{black}{Predictive Control for Driving under Uncertain Road Geometry Estimated from Onboard Vision}}

\author{Jelena Trisovic, Andrea Carron, Melanie N. Zeilinger 
%~\IEEEmembership{affiliations?}
        % <-this % stops a space
\thanks{\color{black} All authors are with the Institute for Dynamic Systems and Control, ETH
Zurich, Switzerland {(email \tt \{tjelena, carrona,  mzeilinger\}@ethz.ch)}.
 Jelena Trisovic is also with ETH AI Center. 
%\newline $^*$ contributed equally to this paper. 
\newline This research was primarily supported by the ETH AI Center through an ETH AI Center doctoral fellowship to Jelena Trisovic. Additionally, it was supported by the Swiss National Science
Foundation under the NCCR Automation (grant 51NF40\_225155).}% <-this % stops a space
}

% The paper headers
%\markboth{Journal of \LaTeX\ Class Files,~Vol.~14, No.~8, August~2021}%
%{Shell \MakeLowercase{\textit{et al.}}: A Sample Article Using IEEEtran.cls for IEEE Journals}

%\IEEEpubid{0000--0000/00\$00.00~\copyright~2021 IEEE}
% Remember, if you use this you must call \IEEEpubidadjcol in the second
% column for its text to clear the IEEEpubid mark.

\maketitle

\begin{abstract}
\textcolor{black} {Autonomous vehicles driving on unknown roads must estimate the road geometry from onboard sensors and follow the resulting reference while respecting road boundaries. When the perception-induced estimation error is non-negligible relative to the lateral constraint margins, treating the estimated reference as ground-truth can lead to safety constraint violations. This paper proposes a perception-based control framework that integrates road geometry estimation and the resulting geometric uncertainty with constrained control.  A parametric model of the road curvature is identified from RGB-D measurements via constrained nonlinear optimization, enforcing geometric consistency.   The residual perception uncertainty is then captured by constructing a set of curvature profiles consistent with measurements by perturbing the estimated parameters. Using the Frenet-frame vehicle model, the curvature uncertainty is propagated through the model via a scenario model predictive control scheme as parametric model uncertainty, thus enforcing constraints across all sampled curvature realizations. The curvature estimation module is evaluated both in high-fidelity simulation and on real-world image data, confirming reliable operation under realistic visual and depth noise. The full perception-to-control pipeline is validated in simulation, demonstrating that the uncertainty-aware controller maintains tighter adherence to road boundaries compared to its nominal counterpart.}
\end{abstract}


\section{Introduction}

\textcolor{black}{Autonomous vehicles operating on roads with no prior geometric map must 
estimate the road geometry from onboard sensors and use it as 
the reference for control. In this context, recent works~\cite{9488179, li2020autonomous} have highlighted the potential of perception-based control methods, which directly operate on representations derived from onboard sensing, to achieve autonomous operation in unseen environments. However, most existing methods implicitly assume perfect 
perception and treat the resulting outputs as certain and  
reliable~\cite{li2020autonomous, betz2022autonomous}.
This decoupled treatment can lead to compounding errors and, 
critically, to safety constraint violations when perception is 
insufficiently accurate.}

\begin{figure}
    \centering
    \includegraphics[width=0.99\linewidth]{ 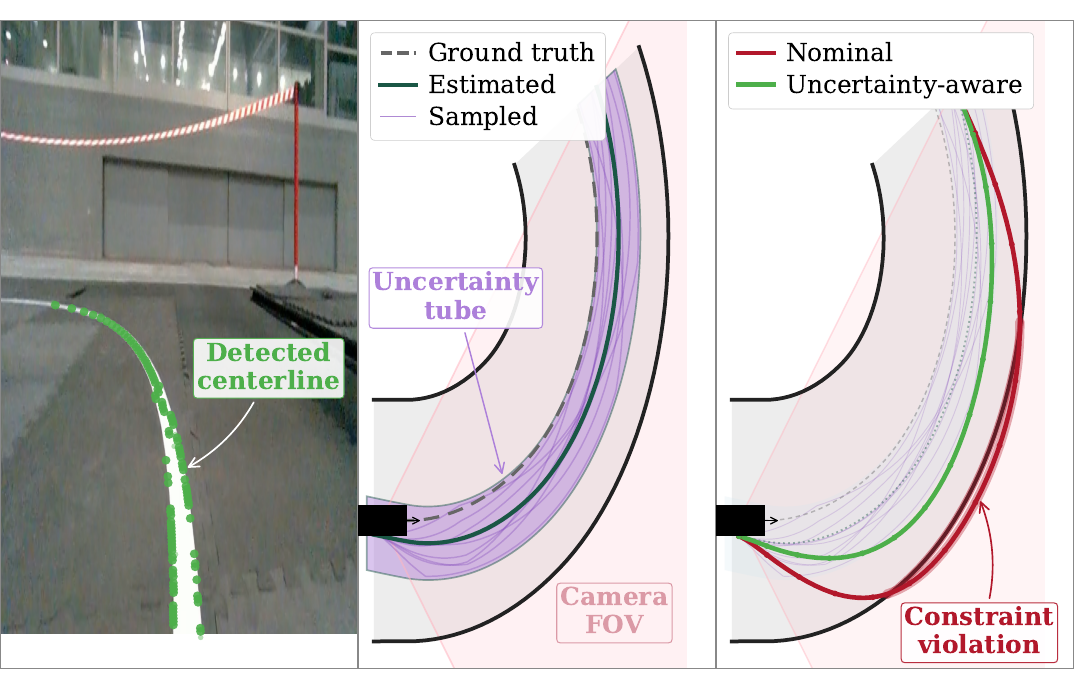}
    \caption{\textcolor{black}{The proposed method estimates the road centerline from onboard RGB-D images (left) and captures perception uncertainty by sampling plausible centerline realizations around the estimate (middle). A controller ignoring this uncertainty violates the true road boundary in closed-loop operation (right, red), while the proposed uncertainty-aware controller ensures safety across all realizations (right, \textcolor{black}{green}).}}
    \label{fig:teaser}
\end{figure}

\textcolor{black}{The challenge of obtaining a reliable road geometry estimate from 
vision is well recognized, with numerous works documenting the 
sensitivity of lane detection to lighting, occlusions, and sensor 
noise~\cite{araujo2024road, waykole2021review}. 
At the same time, an extensive body of work on lane-keeping control 
exists~\cite{kebbati2023lateral}, yet these methods
treat the perception-derived reference as given and only address 
uncertainty in the vehicle dynamics~\cite{bujarbaruah2018adaptive, dahmani2013road, 7893692, 8460385}. When geometric estimation errors are small 
relative to the road width, as is typical on well-marked highways, 
this separation is adequate. In constrained settings, however, such 
as narrow roads, single-lane environments, or small-scale autonomous 
platforms, the estimation error can become comparable to the 
admissible lateral deviation, and certainty-equivalence control may 
fail to satisfy safety constraints. Addressing this limitation requires explicitly incorporating the geometric estimation error into the control formulation.}

\textcolor{black}{A key observation underlying the proposed approach is that 
the Frenet-frame formulation of the vehicle dynamics creates a direct 
 coupling between the road geometry and the system model: the 
curvature of the reference path enters the dynamics as a parameter. When 
the reference is estimated from perception, errors in the estimated road 
geometry manifest as parametric model uncertainty. This 
structural property enables a different treatment of 
perception uncertainty compared to certainty-equivalence approaches.} \textcolor{black}{Building on this insight, we propose a perception-based control framework in which the controller relies on a reference estimated online from camera images and accounts for the resulting uncertainty. Raw images do not enter the \textcolor{black}{controller, instead only} the perception-derived curvature model enters as an uncertain parameter.}

{\color{black} The proposed framework operates in a local, map-free 
setting, relying on locally estimated road geometry to ensure safe 
motion without constructing a global map. To isolate the 
contribution of perception-induced geometric uncertainty, we 
consider a simplified setting in which the vehicle pose is assumed 
known, the road is planar with constant and known width, and the 
environment is static and free of occlusions. The framework is 
agnostic to the specific perception algorithm, as the only required 
interface is a centerline point cloud estimate.}

\begin{figure*}
    \centering
    \includegraphics[width=0.99\linewidth]{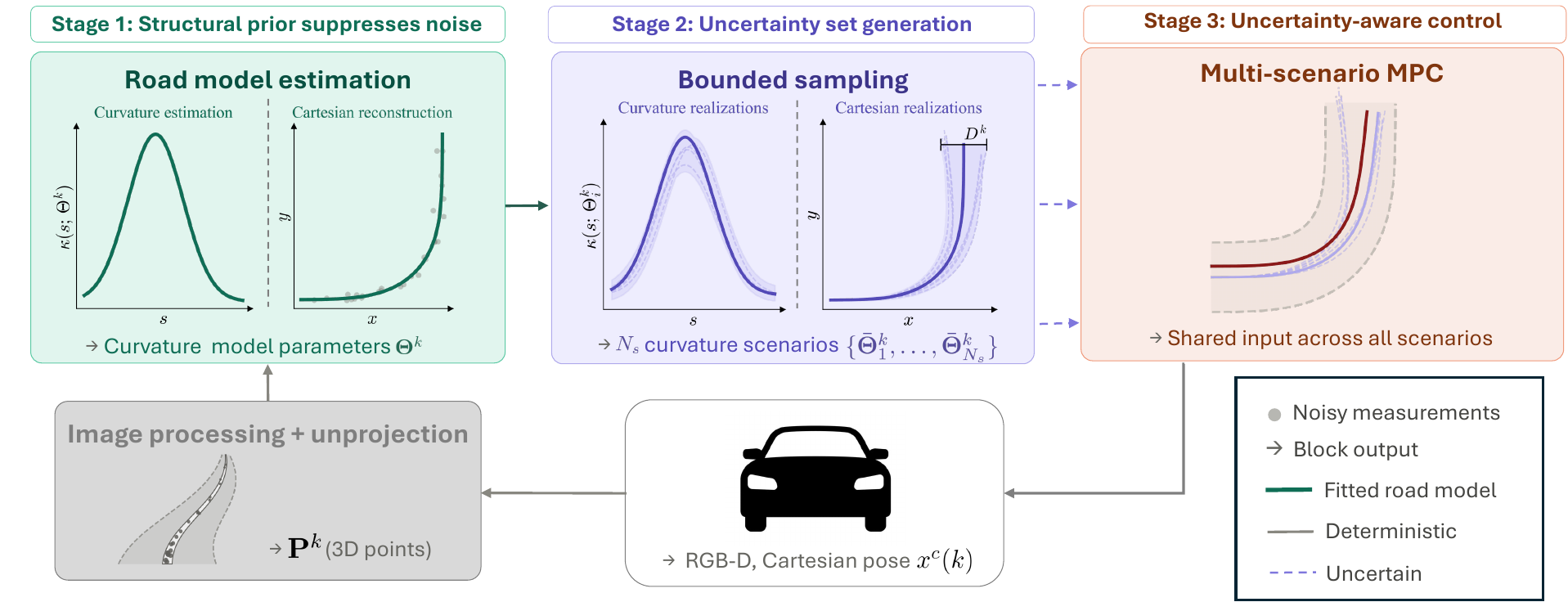}
    \caption{ \textcolor{black}{Overview of the proposed framework.
At each time step $k$, sensor readings (RGB-D images and car pose $x^c(k)$)
are processed to extract an ordered centerline point cloud $\mathbf{P}^k$.
A parametric curvature model $\kappa(s; \mathbf{\Theta}^k)$ is then identified
via constrained nonlinear optimization such that the Cartesian centerline
reconstructed from $\kappa$ fits the extracted noisy \textcolor{black}{data-}points $\mathbf{P}^k$. This
structured estimation acts as a geometric regularizer, attenuating
 perception noise through the imposed curvature prior
(Stage~1). To represent the residual geometric uncertainty, $N_s$ geometrically diverse curvature
realizations $\{\bar{\mathbf{\Theta}}_1^k, \ldots,
\bar{\mathbf{\Theta}}_{N_s}^k\}$ are sampled (Stage~2) such that the reconstructed centerlines lie within a calibrated
Hausdorff-distance $D^k$ from the nominal estimate, defining a bounded ambiguity set. These realizations,
each inducing a distinct set of Frenet-frame dynamics, are passed to a
scenario model predictive controller  that optimizes a shared control input while
enforcing constraint satisfaction across all $N_s$ curvature scenarios.}}
    \label{fig:problem}
\end{figure*}

\textcolor{black}{While the individual components of the proposed 
approach build on established methods, the main contribution lies 
in their integration into a systematic pipeline (Figure~\ref{fig:problem}) \textcolor{black}{ t}hat propagates 
perception uncertainty from estimation to constrained control. The 
specific contributions of this work are:}

{\color{black}
\begin{itemize}
    \item A curvature estimation method that fits a parametric road 
model to a centerline point cloud obtained from any lane marking 
detection algorithm. The model is identified via constrained 
nonlinear optimization that enforces geometric consistency, while attenuating perception noise. By construction, the resulting curvature is twice differentiable and bounded, which guarantees well-posedness of the curvilinear coordinate transformation~\cite{wursching2024robust} and direct compatibility with the optimization-based control formulation.

    \item A sampling-based uncertainty representation that 
    generates multiple plausible road geometries through structured 
    perturbations of the estimated curvature parameters within a 
    calibrated neighborhood reflecting the perception error.
    \item An uncertainty-aware scenario formulation of model 
    predictive contouring control~\cite{5717042, 9802523} that 
    imposes state constraints across all sampled road geometries 
    within a shared control optimization, explicitly integrating 
    perception-induced geometric uncertainty into the control 
    design, as 
    illustrated in Figure~\ref{fig:teaser}.
    
\end{itemize}
}

The remainder of the paper is structured as follows. \textcolor{black}{Section~\ref{sec:rel_work} reviews related literature. Section~\ref{sec:prelims} introduces the problem statement and preliminaries. We give an overview of the proposed method in Section~\ref{sec:overview}, and a detailed description of each module in Sections~\ref{sec:curv_est} (curvature estimation), \ref{sec:uncert_quant} (perception uncertainty representation) and \ref{sec:rob_ctrl} (uncertainty-aware control). The results are presented in Section~\ref{sec:Results}, and Section~\ref{sec:Disc} concludes with a discussion of limitations and future work.}

\textcolor{black}{\textit{Notation:} We denote by $\mathbb{R}$ the set of real numbers, by $\mathbb{R}_{> 0}$ the set of positive real numbers, and by  $\mathbb{Z}_{> 0}$ the set of positive integers. For the set of integers in the interval $[a, b]$, i.e., $\{a, a+1, \dots, b\}$, we use notation $\mathbb{I}_{[a, b]}$. The bold symbol $\mathbf{u}$ refers to a sequence of $n \in \mathbb{Z}_{>0}$ vector-valued variables $u_i \in \mathbb{R}^m$ for $i \in \mathbb{I}_{[1,n]}$, $ \mathbf{u}=[u^{\top}_1, u^{\top}_2, \ldots, u^{\top}_n]^{\top}$. In some cases, to avoid \textcolor{black}{ d}uplicate variable names, we use uppercase bold symbols in place of the lowercase ones. The supremum and infimum of a set are denoted with $\sup$ and $\inf$, respectively, and \textcolor{black}{the} Euclidean norm of vector $x\in$ $\mathbb{R}^n$ is denoted by $\|x\|$.} The notation $\text{diag}(a,b)$ for $a,b \in \mathbb{R}$ is used to represent a diagonal matrix in $\mathbb{R}^{2 \times 2}$ with $a$ and $b$ as its diagonal elements.

\section{Related Work}
\label{sec:rel_work}

{Robots navigating unknown environments rely on perception for situational awareness.} While numerous works \textcolor{black}{a}ddress this challenge across various contexts \cite{8688060,malis2002survey}, they often fail to systematically handle uncertainties and safety throughout the entire autonomy pipeline. \textcolor{black}{In autonomous driving and racing, for example, most methods account for uncertainty in vehicle dynamics and friction~\cite{betz2022autonomous, al2025lla, kalaria2024adaptive},  but assume the road geometry is known a priori, with lidar~\cite{evans2024unifying} and cameras~\cite{kabzan2020amz, baumann2024forzaeth} used primarily for localization rather than road estimation. } Conversely,  the approaches that \textcolor{black}{a}ddress perception uncertainty either do not incorporate realistic dynamical models of the considered autonomous systems (e.g., they assume the robot can stop within a single time step) when integrating them with control \cite{yang2023safe}, or do not evaluate the perception pipeline in closed-loop \cite{pmlrv242mao24c, 10285393}.

In the remainder of this section, we focus on the related \textcolor{black}{work that addresses} the challenge of road curvature estimation and uncertainty-aware control using curvilinear vehicle models, as these topics are central to our approach, which treats the uncertainty in perceived curvature as a model uncertainty. 
\paragraph*{Road curvature estimation} This is a task commonly studied in the context of driver-assistance systems for cars and motorcycles, with the goal of deriving a mathematical description of the road curvature based on visual measurements. The approach introduced in \cite{8938821} performs simultaneous localization and mapping based on estimated road curvature, leveraging its reliability and distinctiveness as a road feature. The method provides precise reconstructions of the map and performs loop closure to mitigate drift issues and ensure global map consistency.  However, it assumes a single value of curvature per perceived image, which is inadequate for roads with rapidly varying curvature, such as racing tracks or mountain roads, where the curvature can change significantly over short distances. The method introduced in \cite{damon2018image} fits a single clothoid (a curvature model inspired by road geometry \cite{casal2017optimization}) to the detected road markers. While similar to our approach, it produces curvature estimates that are neither differentiable nor \textcolor{black}{bounded}, making them unsuitable for optimization-based \textcolor{black}{control} methods. Similarly, in \cite{dahmani2013road} a robust fuzzy observer is used for road curvature estimation, resulting in the same issue with downstream control compatibility. \textcolor{black}{Other works rely on digital road maps~\cite{4621300} or geometric primitives such as Hough-transform-detected lines and splines~\cite{7028086} to estimate the road geometry, while~\cite{7893692} and~\cite{8460385} focus on estimating lateral dynamics under unknown road curvature but consider only its past or present values. To the best of our knowledge, none of these approaches 
jointly address curvature estimation and its downstream 
use in control, nor do they ensure that the estimated 
curvature satisfies the regularity conditions required 
by optimization-based control formulations.}

 \paragraph*{Uncertainty-aware control} \textcolor{black}{Standard robust model predictive control (MPC) formulations~\cite{mayne2005robust, chisci2001systems} ensure constraint satisfaction but require that the uncertainty admits a compact representation and that its effect on the state can be propagated explicitly. In the setting considered here, however, the uncertain road curvature is reconstructed from perception, hence not admitting a priori known bounds, and enters the Frenet-frame dynamics nonlinearly, so neither polytopic nor interval descriptions are readily available.}

\textcolor{black}{Since the Frenet-frame formulation maps geometric uncertainty into parametric model uncertainty, works addressing uncertain curvature in curvilinear vehicle models are most directly related. The approach proposed in \cite{liniger2020safe} formulates a game-theoretic path-following optimization problem where the road curvature acts as an adversary.} Using viability theory, this method computes safe sets as terminal constraints for optimization-based motion planners. \textcolor{black}{While conceptually related, it assumes worst-case curvature over the entire road, resulting in conservative behavior.} {The method introduced in \cite{9762493} robustly addresses model uncertainty in curvilinear MPC by predicting sets of possible trajectories resulting from uncertainties and disturbances. This approach promotes more assertive behavior and tighter constraint adherence early in the optimization horizon, while adopting a more conservative stance toward its end. }\textcolor{black}{Both methods provide rigorous robustness 
guarantees but require either worst-case assumptions over the 
full road geometry or explicit uncertainty sets on the vehicle 
model. In contrast, our method adapts uncertainty to the 
locally estimated road geometry, resulting in a less 
conservative and more context-aware control strategy at the 
cost of not providing formal guarantees.}

\textcolor{black}{\textcolor{black}{Scenario-based approaches, e.g.~\cite{bernardini2011stabilizing, schildbach2014scenario}, avoid explicit set propagation by optimizing a single control sequence subject to feasibility across a finite number of sampled uncertainty realizations, and derive probabilistic constraint satisfaction guarantees from the assumption that scenarios are drawn independently from a common distribution accessible for sampling. A related line of work, multi-stage MPC~\cite{lucia2013multi}, instead considers a finite set of uncertainty realizations within a bounded set. Our formulation adopts the same optimization structure, where multiple curvature realizations are propagated in parallel under a shared control input, and constraints are imposed across all retained scenarios. Since perception-induced curvature uncertainty does not admit a reliable distributional characterization, we follow the multi-stage philosophy and treat it as a bounded geometric deviation in the reconstructed centerline, selecting scenarios to promote geometric diversity within this bounded set. The resulting formulation is therefore a practically motivated instance within this family, rather than a new MPC scheme, and yields empirical robustness with respect to the geometric uncertainty. }}

\section{\textcolor{black}{Problem Statement and Preliminaries}}
\label{sec:prelims}
\label{sec:ps_and_ass}

{\color{black}
We consider an autonomous vehicle with known nonlinear dynamics operating on an unknown planar road of constant and known width~$W$. The vehicle is equipped with an onboard RGB-D camera that provides partial and noisy observations of the road geometry in real time. The vehicle state is assumed to be known, and the road curvature is assumed to be bounded by a known constant $\kappa_{\max}$. We also assume a static environment without dynamic obstacles and neglect occlusions, focusing solely on uncertainty arising from imperfect perception of the road geometry. The objective is to ensure safe motion along the road based solely on locally perceived information. Safety is defined as maintaining the vehicle within the road boundaries at all times, despite uncertainty in the estimated road geometry derived from sensor measurements.

\textcolor{black}{To formalize this problem, we require three 
ingredients. First, a mathematical description of the road 
geometry in terms of its curvature, which serves both as the 
estimation target for the perception module and as the 
reference for control (Section~\ref{subsec:frenet_frame}). Second, 
a vehicle dynamics model expressed in the Frenet frame 
relative to this reference, in which the road curvature 
enters the equations of motion as a parameter 
(Section~\ref{subsec:dyn_mdl}), establishing a direct 
 coupling between the estimated road geometry and 
the system dynamics, and enabling perception uncertainty to be 
treated as parametric model uncertainty. Third, a \textcolor{black}{model predictive contouring control} formulation that imposes 
road boundary constraints while maximizing progress along the 
reference path (Section~\ref{subsec:MPCCF}). The uncertainty-aware 
extension of this controller is presented in 
Section~\ref{sec:rob_ctrl}.}

}

\subsection{\textcolor{black}{Curvature-Based Road Representation}}
\label{subsec:frenet_frame}

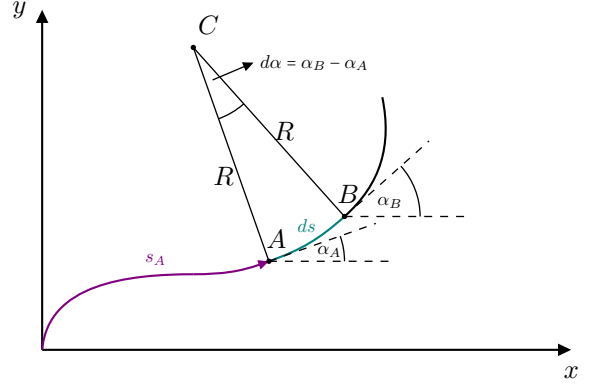
\begin {figure}
\centering
\begin{adjustbox}{max height=0.55\textwidth, max width=0.48\textwidth}
\begin{tikzpicture}[scale=1.0]

    \draw[-{Triangle[length=5pt, width=4.5pt]},line width=0.8pt](0.0, 0.0)--(7.0,0.0);
    \draw[-{Triangle[length=5pt, width=4.5pt]},line width=0.8pt](0.0, 0.0)--(0.0, 4.5);

    \draw[line width=0.8pt, violet] (0.0, 0.0) to [out=85, in=180] (2.0, 1.0);
    \draw[line width=0.8pt, teal] (3, 1.1715729) to [out=19.5, in=41.8-180] (4, 1.7639);
    \draw[-{Triangle[length=4pt, width=3.5pt]},line width=0.8pt, violet] (2.0, 1.0) to [out=0, in=19.5-180] (3, 1.1715729);
    \draw[line width=0.8pt] (4, 1.7639) to [out=41.8, in=56.47-135] (4.5, 3.34168);

    \fill (3, 1.1715729) circle (1pt);   %Point A
    \fill (4, 1.7639)  circle (1pt);     %Point B 
    \fill (2, 4)  circle (1pt);          %Point C

    \draw[line width=0.5pt, dashed] (3, 1.1715729) -- ++(19.5:1.5cm);
    \draw[line width=0.5pt, dashed] (4, 1.7639) -- ++(41.8:1.5cm);
    
    \draw[line width=0.5pt, dashed] (3, 1.1715729) -- ++(0:1.6cm);
    \draw[line width=0.5pt, dashed] (4, 1.7639) -- ++(0:1.6cm);

    \draw[line width=0.5pt] (2, 4) -- (3, 1.1715729);
    \draw[line width=0.5pt] (2, 4) -- (4, 1.7639);

    %%%%% Draw arcs %%%%%

    \draw[line width=0.5pt] (2+0.333644, 3.057) arc[start angle=19.5-90, end angle=41.8-90, radius=1cm];
    \draw[line width=0.5pt] (3+1, 1.1715729) arc[start angle=0.0, end angle=19.5, radius=1cm];
    \draw[line width=0.5pt] (5, 1.7639) arc[start angle=0.0, end angle=41.8, radius=1cm];

    %%%%% Letters %%%%%

    \node[ scale=1.0]() at (-0.3, 4.5) {$y$};
    \node[ scale=1.0]() at (7.0, -0.3) {$x$};
    
    \node[ scale=1.0]() at (2.2, 4.3) {$C$};
    \node[ scale=1.0]() at (3+0.1, 1.1715729+0.3) {$A$};
    \node[ scale=1.0]() at (4+0.05, 1.7639   +0.3) {$B$};

    \node[ scale=0.7](c1) at (3.6, 3.8) {$d\alpha=\alpha_B-\alpha_A$};
    \draw[-{Triangle[length=4pt, width=3.5pt]},line width=0.5pt](2.2548, 3.5697)-- (c1.west);
    \node[ scale=0.7](c2) at (3.8, 1.3) {$\alpha_A$};
    \node[ scale=0.7](c3) at (4.6, 1.95) {$\alpha_B$};

    \node[ scale=0.7, violet](c4) at (1.5, 1.2) {$s_A$};
    \node[ scale=0.7, teal](c5) at (3.5, 1.65) {$ds$};

    \node[ scale=1.0](c6) at (2.4, 2.3) {$R$};
    \node[ scale=1.0](c7) at (3.2, 2.9) {$R$};

    %%%%%%

\end{tikzpicture}
\end{adjustbox}
    \caption{\textcolor{black}{Coordinates of points on a curve. Arc length and tangent angle of point $A$ are denoted by $s_A$ and $\alpha_A$. Point C denotes the center of the arc with radius $R$, limited by points $A$ and $B$.} Given the relationship between the coordinates of points $A$ and $B$, equation~\eqref{eq:forw_prop1} can be derived for $ds \rightarrow 0$.} 
\label{fig:img3}
\end{figure}

In this subsection, we review how a curve can be described in both Frenet and Cartesian frames \cite{dickmanns1987curvature}. Consider a curve $\gamma: \mathbb{R} \rightarrow \mathbb{R}^2$,  $\gamma(s)=\left[x(s), y(s)\right]^{\top}$ (as shown in Figure \ref{fig:img3}), where $s$ is its arc length, and $\alpha(s)$ denotes the tangent angle \textcolor{black}{at} each point of the curve. The curve $\gamma(s)$ can then be described by its initial tangent angle $\alpha_0$, and curvature $\kappa(s)=\frac{\mathrm{d} \alpha(s)}{\mathrm{d} s}$. 

The Cartesian coordinates of the curve, $\left[x(s), y(s)\right]^{\top}$, can be reconstructed from the curvature $\kappa(s)$, initial curve coordinates in Cartesian frame, $(x_0,y_0)$, and \textcolor{black}{initial} tangent angle $\alpha_0$ using the following relationship

\begin{equation}
\begin{aligned}
    \alpha(s) &= \alpha_0 + \int_{0}^{s} \kappa(\lambda) \,d\lambda,\\ 
    x(s) &= x_0 + \int_{0}^{s} \cos(\alpha(\lambda))\,d\lambda, \\
    y(s) &= y_0 + \int_{0}^{s} \sin(\alpha(\lambda))\,d\lambda,
\end{aligned}
    \label{eq:forw_prop1}
\end{equation}
\textcolor{black}{where $\lambda$ is used as an integration variable for curvature estimation of the road model, while $s$ denotes the arc length in the curve definition and vehicle's progress variable in the Frenet frame dynamics and MPC.}

\textcolor{black}{To obtain the Cartesian coordinates from curvature, we numerically integrate equation~\eqref{eq:forw_prop1}, and denote its discretization with $c_{k+1}=f_{\Delta}(c_k, \kappa(s_k))$, where $c_k=\left[\alpha_k, x_k, y_k \right]^\top$ represents the corresponding discretized values of  points along the curve with curvature $\kappa(s)$ at arc length $s=s_k$. }

\subsection{Bicycle Model of a Car in the Frenet Frame} 
\label{subsec:dyn_mdl}

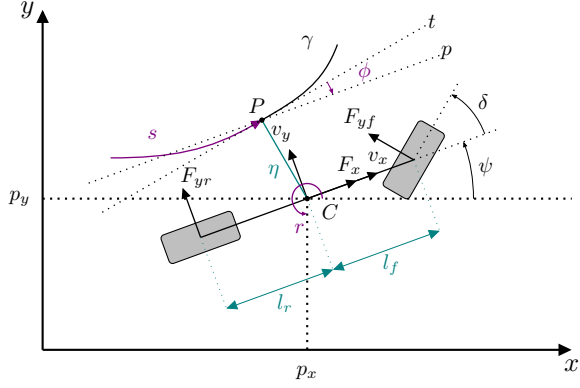
\begin {figure}
\centering
\begin{adjustbox}{max height=0.55\textwidth, max width=0.48\textwidth}
\begin{tikzpicture}[scale=1.0]

    \draw[line width=0.8pt, dotted] (-3.50, 0.0) to  (3.50, 0.0);
    
    \coordinate (P1) at ($(0,0) + (20:1.5cm)$);
    \coordinate (P2) at ($(0,0) + (200:1.5cm)$);
    \coordinate (C) at (0,0);

    \node[draw, rectangle, rounded corners=0.5mm, minimum width=1cm, minimum height=0.4cm, anchor=center, fill=lightgray, rotate=20] at (P2) {};
    \node[draw, rectangle, rounded corners=0.5mm, minimum width=1cm, minimum height=0.4cm, anchor=center, fill=lightgray, rotate=60] at (P1) {};    

    \draw[line width=0.5pt] (0,0) -- ++(20:1.5cm);
    \draw[line width=0.5pt] (0,0) -- ++(200:1.5cm);
    \draw[line width=0.8pt, dotted] (0,0) -- (0.0, -2);

    \coordinate (XC) at (0,-2.3);
    \coordinate (YC) at (-3.8,0);
    \node[ scale=0.8, black]() at (XC)  {$p_x$};
    \node[ scale=0.8, black]() at (YC)  {$p_y$};

    %%%%% Angle lines %%%%%

    \draw[line width=0.5pt, dotted] (P1) -- ++(20:1.2cm);
    \draw[line width=0.5pt, dotted] (P1) -- ++(60:1.2cm);
    \draw[-{Triangle[length=4pt, width=3.5pt]},line width=0.5pt] (P1) -- ++(150:0.7cm);

    \draw[-{Triangle[length=2pt, width=1.5pt]}, line width=0.3pt] (P1) ++(20:1cm) arc[start angle=20, end angle=60, radius=1cm];
    \draw[-{Triangle[length=2pt, width=1.5pt]}, line width=0.3pt] (C) ++(0:2.2cm) arc[start angle=0, end angle=20, radius=2.2cm];

    \draw[-{Triangle[length=4pt, width=3.5pt]},line width=0.5pt] (C) -- ++(20:0.7cm);
    \draw[-{Triangle[length=4pt, width=3.5pt]},line width=0.5pt] (C) ++(20:0.7cm) -- ++(20:0.3cm);
    
    \draw[-{Triangle[length=4pt, width=3.5pt]},line width=0.5pt] (C) -- ++(110:0.7cm);
    \draw[-{Triangle[length=4pt, width=3.5pt]},line width=0.5pt] (P2) -- ++(110:0.7cm);

    %%%%% Cote %%%%%

    \coordinate (P3) at ($(0,0) + (290:1cm)$);
    \coordinate (P4) at ($(P3)  + (20:1.5cm)$);
    \coordinate (P5) at ($(P3)  + (200:1.5cm)$);    

    \draw[line width=0.3pt, dotted, teal] (C) -- (P3);
    \draw[line width=0.3pt, dotted, teal] (P1) -- (P4);
    \draw[line width=0.3pt, dotted, teal] (P2) -- (P5);

    \draw[{Triangle[length=4pt, width=3.5pt]}-{Triangle[length=4pt, width=3.5pt]}, line width=0.3pt, teal] (P5) -- (P3);
    \draw[{Triangle[length=4pt, width=3.5pt]}-{Triangle[length=4pt, width=3.5pt]}, line width=0.3pt, teal] (P3) -- (P4);

    %%%%% Road %%%%%

    \coordinate (P6) at ($(0,0) + (120:1.2cm)$);
    \coordinate (P7) at ($(P6) + (30:1.2cm)$);
    \coordinate (P8) at ($(P6) + (210:1.2cm)$);

    \coordinate (PP) at ($(P6) + (110:0.2cm)$);
    \coordinate (PC) at ($(C) + (330:0.35cm)$);
    \node[ scale=0.8, black]() at (PP)  {$P$};
    \node[ scale=0.8, black]() at (PC)  {$C$};

    \coordinate (P_7) at ($(P6) + (30:2.5cm)$);
    \coordinate (P_8) at ($(P6) + (210:2.5cm)$);

    \draw[line width=0.5pt, teal] (C) -- (P6);

    \coordinate (P9) at ($(P6) + (180:2cm) + (270:0.5cm)$);
    \coordinate (P10) at ($(P6) + (0:1cm) + (90:1cm)$);

    \draw[-{Triangle[length=4pt, width=3.5pt]},line width=0.5pt, violet  ] (P9) to [out=0, in=210] (P6);
    \draw[line width=0.5pt] (P6) to [out=30, in=250] (P10);
    
    \draw[-{Triangle[length=2pt, width=1.5pt]},line width=0.5pt, violet  ] (C)  ++(0:0.2cm) arc[start angle=0, end angle=270, radius=0.2cm];
    
    \draw[line width=0.5pt, dotted] (P_7) -- (P_8);  %tangent to the curve

    \coordinate (PP7) at ($(C) + (30:1.2cm)$);
    \coordinate (PP8) at ($(C) + (210:1.2cm)$);

    \coordinate (PPP7) at ($(P6) + (20:2.5cm)$);
    \coordinate (PPP8) at ($(P6) + (200:2.5cm)$);

    \coordinate (PPP9) at ($(P6) + (20:2.6cm)$);
    \coordinate (PPP10) at ($(P6) + (30:2.6cm)$);
    \node[ scale=0.8, black]() at (PPP9)  {$p$};
    \node[ scale=0.8, black]() at (PPP10)  {$t$};

    \draw[-{Triangle[length=2pt, width=1.5pt]},line width=0.3pt, violet] (P6) ++(30:1cm) arc[start angle=
    30, end angle=20, radius=1cm];
    
    \coordinate (PPP11) at ($(P6) + (25:1.5cm)$);
    \node[ scale=0.8, violet]() at (PPP11)  {$\phi$};

    \draw[line width=0.5pt, dotted] (PPP7) -- (PPP8);  %tangent to the curve

    %%%%% Letters %%%%%

    \coordinate (L1) at ($(P3)!0.5!(P4)$);
    \coordinate (L2) at ($(P3)!0.5!(P5)$);
    \coordinate (L11) at ($(L1)+(290:0.2cm)$);
    \coordinate (L21) at ($(L2)+(290:0.2cm)$);
    
    \node[ scale=0.8, teal]() at (L11)  {$l_f$};
    \node[ scale=0.8, teal]() at (L21)  {$l_r$};

    \coordinate (L3) at ($(C) +(10:2.4cm)$);
    \coordinate (L4) at ($(P1)+(40:1.2cm)$);
    \node[ scale=0.8]() at (L3)  {$\psi$};
    \node[ scale=0.8]() at (L4)  {$\delta$};

    \coordinate (L5) at ($(C)  + (20:0.7cm)$);
    \coordinate (L6) at ($(C)  + (20:1.0cm)$);
    \coordinate (L7) at ($(P1) + (150:0.7cm)$);
    \coordinate (L8) at ($(C)  + (110:0.7cm)$);
    \coordinate (L9) at ($(P2) + (110:0.7cm)$);

    \coordinate (L51) at ($(L5)  + (90:0.2cm) + (180:0.1cm)$);
    \coordinate (L61) at ($(L6)  + (90:0.2cm)$);
    \coordinate (L71) at ($(L7)  + (90:0.2cm)+ (180:0.1cm)$);
    \coordinate (L81) at ($(L8)  + (90:0.2cm)+ (180:0.1cm)$);
    \coordinate (L91) at ($(L9)  + (90:0.2cm)+ (180:0.1cm)$);
    \coordinate (L9_1) at ($(L9)  + (90:0.2cm)+ (340:0.2cm)$);
    \coordinate (L101) at ($(L7)  + (90:1.2cm)+ (180:0.8cm)$);
    
    \node[ scale=0.8]() at (L51)  {$F_x$};
    \node[ scale=0.8]() at (L61)  {$v_x$};
    \node[ scale=0.8]() at (L71)  {$F_{yf}$};
    \node[ scale=0.8]() at (L81)  {$v_y$};
    \node[ scale=0.8]() at (L9_1)  {$F_{yr}$};
    \node[ scale=0.8, black]() at (L101)  {$\gamma$};

    \coordinate (L10) at ($(C)  + (270:0.4cm) + (180:0.1cm)$);
    \node[ scale=0.8, violet]() at (L10)  {$r$};

    \coordinate (L11) at ($(L91)  + (90:0.5cm) + (0:-0.3cm)$);
    \node[ scale=0.8, violet]() at (L11)  {$s$};

    \coordinate (L12) at ($(L81)  + (270:0.5cm) + (180:0.1cm)$);
    \node[ scale=0.8, teal]() at (L12)  {$\eta$};

    %%%%% Coord syst %%%%%

    \draw[-{Triangle[length=5pt, width=4.5pt]},line width=0.8pt](-3.5, -2)--(3.5,-2);
    \draw[-{Triangle[length=5pt, width=4.5pt]},line width=0.8pt](-3.5, -2)--(-3.5, 2.5);
    \node[ scale=1]() at (3.5,-2.2)  {$x$};
    \node[ scale=1]() at (-3.7, 2.5)   {$y$};

    \fill (P6) circle (1pt);
    \fill (0,0) circle (1pt);
    
\end{tikzpicture}
\end{adjustbox}
\caption{{Car state in the Frenet frame with respect to a reference path $\gamma$. The figure also shows relevant physical forces and control variables.}}
\label{fig:img4}
\end{figure}

{The state of the car in Frenet frame is given by the vector $x=\left[s, \eta, \phi, v_x, v_y, r, \delta, \tau \right]^\top$, where $s, \eta,$ and $\phi$ denote the progress, lateral distance and heading with respect to the reference path denoted by $\gamma $, as illustrated in Figure~\ref{fig:img4}. The car's center of gravity is denoted by $C$ and its projection onto $\gamma$ with $P$. The tangent to $\gamma$ in $P$ is denoted with $t$, and the line parallel to the axis of the car that passes through $P$ with $p$. The progress $s$ of the car is defined as the arc length of the section of $\gamma$ between its initial point and $P$. The lateral distance $\eta$ is equal to the  distance between $C$ and $P$, while the heading angle $\phi$ is defined as the directed angle between $t$ and $p$. Longitudinal and lateral velocities, $v_x$ and $v_y$, are defined along the car axis and its normal, respectively, while $r$ denotes the angular velocity. The steering angle $\delta$ is defined as the orientation of the front wheel and $\tau$ is the drivetrain input. \textcolor{black}{The Cartesian state of the car is given by $x^c=\left[p_x, p_y, \psi, v_x, v_y, r, \delta, \tau \right]^\top$, where $p_x, p_y$ and $\psi$ correspond to the Cartesian coordinates of the center of mass and the yaw angle of the car. Note that the Cartesian and Frenet state of the car only differ in the first three dimensions, while the remaining ones are shared.}  
}

The car dynamics in the Frenet frame is given by 

\begin{equation}
    \dot{x}=\left[\begin{array}{c}
 \frac{v_x \cos (\phi)-v_y \sin (\phi)}{1-\eta\kappa(s)} \\
v_x \sin (\phi)+v_y \cos (\phi) \\
r-\kappa(s) \frac{v_x \cos (\phi)-v_y \sin (\phi)}{1-\eta\kappa(s)} \\
\frac{1}{m}\left(F_x-F_{y f} \sin (\delta)+m v_y r\right) \\
\frac{1}{m}\left(F_{y r}+F_{y f} \cos (\delta)-m v_x r\right) \\
\frac{1}{I_z}\left(F_{y f} l_f \cos (\delta)-F_{y r} l_r\right) \\
\dot{\delta} \\
\dot{\tau}
\end{array}\right],
\label{eq:curv_dyn}
\end{equation}
where $m$ is the car mass, $I_z$ is its yaw moment of inertia, and $l_{f / r}$ is the distance between its center of gravity and the front and rear axles, respectively. Importantly, note that the curvature of the reference trajectory, $\kappa(s)$, also enters the model. While the physical inputs to the system are $\delta$ and $\tau$, the input applied to the given model is defined as the rate of change in steering angle and drivetrain, i.e., as $u=[\dot{\delta}, \dot{\tau}]^{\top}$. This formulation allows for a more accurate representation of actuator dynamics and simplifies the enforcement of input rate constraints, effectively preventing abrupt control behavior. 

The lateral tire forces $F_{y f}$ and $F_{y r}$ are modeled using a simplified Pacejka tire model \cite{pacejka2005tire}
$$
\begin{aligned}
& \alpha_f=\arctan \left(\frac{v_y+l_f r}{v_x}\right)-\delta, \alpha_r=\arctan \left(\frac{v_y-l_r r}{v_x}\right), \\
& F_{y f / y r}=D_{f / r} \sin \left(C_{f / r} \arctan \left(B_{f / r} \alpha_{f / r}\right)\right),
\end{aligned}
$$
\\
where $\alpha_f$ and $\alpha_r$ are the tire slip angles and $B_{f/r}, C_{f/r}$ and $D_{f/r}$ are the model constants. The longitudinal force is modeled as a single force applied to the center of gravity of the vehicle and is computed as  $F_x=C_1 \tau+$ $C_2 \tau^2+C_3 v_x+C_4 v_x^2+C_5 \tau v_x + C_6$ for some constants $C_1, C_2, C_3, C_4, C_5$ and $C_6$. The drivetrain input $\tau$ can be positive, resulting in forward motion, or negative, resulting in braking. For the remainder of this paper, we denote the discretization of equation~\eqref{eq:curv_dyn} as  $x_{i+1}=f^d(x_i,u_i, \kappa(\cdot))$.

\subsection{Model Predictive Contouring Control in Frenet Frame}
\label{subsec:MPCCF}
In this section, we present the  model predictive contouring control (MPCC) framework, following a formulation similar to that in \cite{vazquez2020optimization}. The primary objective of MPCC is to maximize the vehicle's progress along the reference path while ensuring compliance with both the road boundaries and the car’s dynamics. 

{Let} $N\in\mathbb{Z}_{>0}$ denote the horizon length, $x(k)\in \mathbb{R}^8$ denote the system state in Frenet frame at time $k$, and $\mathbf{X}=\left[{{x}}^{\top}_0, \ldots, {{x}}^{\top}_N\right]^{\top} \text { and } \mathbf{U}=\left[{u}^{\top}_0, \ldots, {u}^{\top}_{N-1}\right]^{\top}$ correspond to the sequences of the system states and inputs over the horizon. {The} control problem can be formulated as the following optimization problem

\begin{equation}
    \begin{aligned}
    \min_{\mathbf{X}, \mathbf{U}} & \sum_{i = 0}^{N-1}l_i(x_i, u_i) + L_N(x_N)
\\ 
\text{s.t. } & x_0 = x(k) \\
& x_{i+1} = f^d(x_i,u_i, \kappa(\cdot)) \\ 
& \textcolor{black}{x_i \in \mathcal{X}, u_i \in \mathcal{U},  x_N \in \mathcal{X}_f}.
 \end{aligned}
\label{eq:curv_dynamics}
\end{equation}
State and input constraints are denoted by $\mathcal{X}$ 
and $\mathcal{U}$, respectively, and $\mathcal{X}_f$ denotes 
an optional terminal constraint set. In the nominal 
formulation, no terminal constraints beyond the state 
constraints are imposed, i.e., $\mathcal{X}_f = \mathcal{X}$. 
Within $\mathcal{X}$, track constraints can be formulated as  $ -\frac{W}{2} \leq \eta_i \leq \frac{W}{2}$ (where $W$ is road width) and $ \phi_{min} \leq \phi_i \leq \phi_{max}$, ensuring that the car stays within a specified lateral distance and keeps within the allowed range of orientations $[ \phi_{min},  \phi_{max}]$ with respect to the centerline. Additional box constraints are imposed on system states and inputs to ensure that the physical limitations of the system are taken into account. 

For some positive constants $q_s, q_{\eta}, q_{\phi}$ and a positive definite matrix $R$, we define the stage cost as 
\begin{equation*} 
    l_i({x}_i, {u}_i){=}
    \begin{cases}
     \textcolor{black}{q_{\eta} \eta_i^2{+}q_\phi \phi_i^2{+}{u_i}^T R {u_i},} & i\in \mathbb{I}_{[1,N-1]} \\
     \textcolor{black}{-q_s s_{i}{+}q_{\eta} \eta_i^2{+}q_\phi \phi_i^2,} & \textcolor{black}{i=N} \\
      {u_0}^T R {u_0}, & i{=}0.
    \end{cases}
  \end{equation*}
The goal of this cost is to incentivize \textcolor{black}{p}rogress along the reference path, while keeping the lateral and orientation errors low. For some positive constants $q_{vx}$ and $q_{vy}$, the terminal cost is chosen as
$L_N=q_{vx}v_{x,i}^2+q_{vy}v_{y,i}^2$ to encourage low velocities at the end of the horizon. 
The optimization problem is solved in a receding horizon fashion, i.e., at each time step $k$ an optimal control sequence $\mathbf{U}^*$ is obtained, but only the first element of this sequence, $u_0^*$, is  applied to the system.

\textcolor{black}{\textcolor{black}{With these elements in place, the considered 
problem can be formulated more precisely. At each time step 
$k$, the perception module provides a noisy centerline point 
cloud $\mathbf{P}^k$ from which the curvature $\kappa(s)$ 
is estimated. Since this estimate is uncertain, the 
Frenet-frame dynamics~\eqref{eq:curv_dyn} operate under 
parametric uncertainty. The objective is to determine a 
control input $u_k$ that satisfies the road boundary 
constraints $|\eta_i| \leq \tfrac{W}{2}$ not only for the 
nominal centerline estimate, but for all plausible road 
geometries consistent with the perception accuracy.}}

\section{\textcolor{black}{Method Overview}}
\label{sec:overview}
{\color{black}

The proposed framework, illustrated in Fig.~\ref{fig:problem}, consists of a perception module and a control module. At each time step $k$, the perception module processes RGB-D measurements and the vehicle state to produce a finite set of $N_s$ road models 
${\overline{\mathbf{\Theta}}_1^k}, {\overline{\mathbf{\Theta}}_2^k}, \cdots , {\overline{\mathbf{\Theta}}_{N_s}^k}$
that represent plausible realizations of the road geometry. The control module incorporates this set into a scenario MPC problem, which imposes constraint satisfaction with respect to all realizations.

The perception module is composed of three submodules: (i) image processing, (ii) curvature estimation, and (iii) sampling. First, pixels corresponding to the road centerline are extracted from the RGB image using any centerline detection method, e.g.,~\cite{8938821}. Their corresponding 3D coordinates in the camera frame are obtained via unprojection using the camera parameters, depth measurements, and pixel locations~\cite{hartley2003multiple}. These points are then transformed into 2D world frame coordinates (under the assumption that the road is flat) using the current vehicle pose given by the Cartesian state $x^c(k)$. The resulting ordered set of $M_k$ two-dimensional centerline points $\mathbf{P}^k = \{p_i^k\}_{i=1}^{M_k}=\{\left[p_{xi}^k, p_{yi}^k\right]^\top\}_{i=1}^{M_k}$ is inherently noisy due to sensing errors (e.g., lighting or depth noise). However, analytical propagation of pixel-level noise through the full perception pipeline is intractable, and standard noise models (Gaussian or bounded) do not adequately capture the resulting geometric uncertainty.  Rather than modeling noise at the measurement level, the proposed formulation therefore abstracts all perception-induced effects into a bounded set of geometrically plausible centerlines.

\textcolor{black}{A key consideration in the choice of 
uncertainty representation is that curvature is a 
second-order differential property of the centerline. The 
mapping from curvature error to Cartesian reconstruction 
error is therefore nonlinear: large local curvature 
deviations may induce negligible positional offsets, while 
small sustained biases can accumulate into significant 
lateral errors. Characterizing uncertainty directly in 
curvature space, whether through distributional assumptions 
or worst-case bounds, therefore yields a representation 
that either does not directly capture the actual geometric 
error or is overly conservative. For this reason, we 
represent perception-induced uncertainty as a bounded 
discrepancy at the level of the reconstructed centerline, 
which is the quantity directly enforcing lateral constraint 
satisfaction.}

To construct this set, we proceed in two steps. First, we fit a parametric road model to the noisy point cloud, producing a nominal centerline estimate described by curvature parameters $\mathbf{\Theta}^k$ (Section~\ref{sec:curv_est}). Second, we construct a bounded neighborhood around this estimate that captures the residual geometric uncertainty due to imperfect model fitting or insufficient modeling expressiveness (Section~\ref{sec:uncert_quant}). The sampled centerlines within this neighborhood are then embedded into the vehicle dynamics via the Frenet formulation, where curvature enters as a model parameter, and incorporated into the scenario MPC (Section~\ref{sec:rob_ctrl}), which imposes constraint satisfaction across all centerline realizations.
}

\section{Curvature Estimation}
\label{sec:curv_est}

{\color{black}

Given the centerline point cloud $\mathbf{P}^k$, the proposed estimation pipeline fits 
curvature parameters $\mathbf{\Theta}^k$ defining a nominal 
centerline model, and is structured as follows. In 
Section~\ref{subsec:roadmodel}, we introduce the parametric road 
model, which represents curvature as a sum of sigmoid functions. 
This model encodes prior knowledge about road geometry and enforces 
physical consistency, thereby attenuating perception 
artifacts. It provides a differentiable road representation with bounded 
curvature, suitable for subsequent uncertainty sampling and 
optimization-based control. Given this model, the curvature 
parameters are estimated from measurements. When no prior estimate 
is available (i.e., at $k=0$), the parameters are obtained using 
the optimization procedure described in 
Section~\ref{subsec:init_est}. This step fits the parametric model to the measured point cloud $\mathbf{P}^k$ by minimizing the discrepancy in Cartesian coordinates between the measured points and their reconstructed counterparts under the fitted road model, subject to geometric constraints. For subsequent time steps 
($k > 0$), new measurements are incorporated through the update 
procedure described in Section~\ref{subsec:est}. This step refines 
the existing model by combining previously estimated geometry with 
newly observed points, ensuring temporal consistency while adapting 
to newly perceived road segments. The complete curvature estimation 
procedure is summarized in Section~\ref{subsec:algo}.

}

\subsection{Road Curvature Model}
\label{subsec:roadmodel}

{\color{black} Road curvature is piecewise-linear by design~\cite{casal2017optimization}, with constant-curvature segments (e.g., straights and circular arcs) connected by linear transitions. To capture this, we model curvature as a sum of sigmoid functions, each representing a differentiable approximation of a single curvature transition.  

The sigmoid parametrization is particularly well-suited for integration with the proposed control framework. Since sigmoids are infinitely differentiable, unlike 
piecewise-linear representations, which require additional 
continuity conditions and are not directly amenable to 
gradient-based optimization, the resulting curvature profile 
inherits the second-order differentiability and bounded curvature required for 
the well-posedness of the curvilinear coordinate 
transformation~\cite{wursching2024robust} and for 
compatibility with contouring 
MPC~\cite{5717042, 9802523}. Moreover, each sigmoid saturates away from its transition center, so its influence on the curvature profile is spatially localized. Once a curve segment lies behind the vehicle, the corresponding sigmoid reduces to a constant offset and can be discarded without affecting the active portion of the model. The parametrization is also low-dimensional, as the number of parameters scales with the number of curvature transitions rather than with the spatial resolution of the centerline. Additionally, their admissible ranges follow directly from road design specifications (e.g., maximum curvature, maximum arc length), which enables explicit imposing of geometric constraints during estimation and facilitates direct sampling in the parameter space for uncertainty propagation (Section~\ref{sec:uncert_quant}). Alternative representations such as Gaussian processes or
splines could in principle also provide differentiable curvature profiles, but
enforcing hard bounds on the function value and its derivatives within
these frameworks is nontrivial and would sacrifice the low-dimensional,
directly constrained parametrization that the sigmoid model offers.
}

We define curvature $\kappa$ as a function of arc length $s$ and express it as a sum of sigmoid functions, given by
\begin{equation*}
    \sigma_i(s; \theta_i)=\frac{a_i}{1+e^{-c_i(s-b_i)}},
\end{equation*}
where $\theta_i=[a_i, b_i, c_i]^\top \in\mathbb{R}^3 $ are the parameters of the sigmoid function. Each parameter has a clear interpretation: $a_i$ indicates the change in curvature magnitude, $b_i$ represents the transition point between two linear sections, and $c_i$ controls the steepness of the transition. Larger values of $c_i$ result in sharper transitions, effectively approximating piecewise-constant functions, while smaller values yield smoother transitions, suitable for modeling piecewise-linear curvature. \textcolor{black}{ The curvature at any arc length $s$ is then expressed as the sum of $n$ sigmoid functions, with an additional offset  $\kappa_0 \in \mathbb{R}$}:

\begin{equation*}
    \kappa(s; \mathbf{\Theta})=\kappa_0 + \sum_{i = 1}^n \sigma_i(s; \theta_i),
\end{equation*}
where $\mathbf{\Theta}$ is the vector of all sigmoid parameters, i.e., $\mathbf{\Theta}=~\left[ \kappa_0, \theta_1^\top, \theta_2^\top, \dots ,\theta_n ^\top\right]^\top$. 

{
\color{black}
We reconstruct Cartesian centerline coordinates from the
curvature profile $\kappa(s;\mathbf{\Theta})$ via a discrete
approximation of~\eqref{eq:forw_prop1}. We denote the $i$-th point at time step $k$ by $m_i^k$, and the resulting ordered set of $L_k$ points (i.e., the map) by $\mathbf{m}^k = \{m_i^k\}_{i=1}^{L_k}$. In addition to position, we associate each point with the corresponding tangential angle $\alpha_i^k$ of the centerline. We define the extended map vector at point $i$ as ${\mu_{i}^{k}} = [\alpha_{i}^k, {(m_{i}^{k})}^{\top}]^{\top}$ and collect all such vectors into the sequence $\boldsymbol{\mu}_k = \{\mu_i^k\}_{i=1}^{L_k}$.

The proposed curvature representation is defined over a finite spatial horizon ahead of the vehicle. Both the horizon length and the maximum number of sigmoids are design parameters chosen based on the road scale, sensor field of view, and expected vehicle speed. The mechanism by which this finite representation is maintained as the vehicle progresses is described in Section~\ref{subsec:est}.
}

\subsection{Initial Curvature Estimation}
\label{subsec:init_est}

\textcolor{black}{In this subsection, we describe how the road curvature model from Section \ref{subsec:roadmodel} is estimated using $M_0$ centerline point coordinates $\mathbf{P}^0 = \{{p_i^0}\}_{i=1}^{M_0}$ extracted from the first RGB-D measurement (i.e., at time step $k=0$). As $p_i^0$ are noisy observations, they do not necessarily satisfy the structural assumptions of the road model. Therefore,  our goal is to estimate the curvature parameters $\mathbf{\Theta}^0\in \mathbb{R}^{3n_0+1}$  of $n_0 \in \mathbb{Z}_{>0}$ sigmoids, the initial tangential angle $\alpha_0^0\in \mathbb{R}$ and the map points $\mathbf{m}^0 \in~\mathbb{R}^{2 \times M_0}$, such that the points $\mathbf{m}^0$ are constrained to lie on the parametric curvature-based road model described in Section~\ref{subsec:roadmodel}, while approximating the measurements $\mathbf{P}^0$. For $\mathbf{Z}^0=[\alpha_0^0,  {\mathbf{\Theta}^{0}}^{\top}, {\mathbf{m}^{0}}^{\top} ]^{\top}$, this is formulated as the following optimization problem }

\begin{equation}
\begin{aligned}
    \min_{\mathbf{Z}^0} & \sum_{i = 0}^{M_0-1}{\lVert {p_{i}^0} - {{m}_{i}^{0}} \lVert}^2 + \mathcal{L}(\mathbf{\Theta}^0)
\\
\text{s.t. } & \textcolor{black}{{m_0^{0}}={p_0^0},  \lambda_0 = 0 }\\
& \lambda_{i+1} = \lambda_i + \Delta \lambda_i \\ 
&  \begin{bmatrix}
           \alpha_{i+1}^0 \\
           {m_{i+1}^{0}}
         \end{bmatrix} = f_{\Delta}\left(\begin{bmatrix}
           \alpha_{i}^0 \\
           {m_{i}^{0}}
         \end{bmatrix}, \kappa(\lambda_i; \mathbf{\Theta}^0) \right)\\
& \textcolor{black}{\alpha_0^0 \in (-\pi, \pi], {m_i^{0}} \in \mathcal{P}, \mathbf{\Theta}^0 \in \mathcal{Q},} \\
\end{aligned}
\label{eq:map_init}
\end{equation}
\textcolor{black}{where $\Delta \lambda_i$ denotes arc length between $p_i^0$ and $p_{i+1}^0$, which is approximated by their Euclidean distance due to their close proximity. Variable $\lambda_i$ denotes the integration parameter representing equally spaced points along the arc length.} \textcolor{black}{Additionally, $f_{\Delta}(\cdot)$ is the discretized curve integration function given by equation~\eqref{eq:forw_prop1}, and $\mathcal{L}(\cdot)$ represents {an optional} curvature parameter regularization. The number of sigmoids,  $n_0$, is either inferred during initialization from the detected curvature transitions, or fixed as a hyperparameter.} The constraints on the reconstructed points and curvature parameters are denoted by $\mathcal{P}$ and $\mathcal{Q}$, respectively. Specifically, for given curvature bounds $\underline{\kappa}$ and $\overline{\kappa}$,  $\mathcal{Q}$ includes the following constraints:
\begin{subequations}\label{eq:constrs}
\begin{align}
     & \textcolor{black}{0 \leq b_i < b_j \leq \lambda_{M_0} \text{ for } i < j}  \label{eq:constr_ordering}\\ 
    % & b_i \geq 0 \label{eq:sconstr_positive}\\
    % & \textcolor{black}{b_{n_0} \leq \lambda_{M_0} \label{eq:constr_latest}}\\
    & \underline{\kappa} \leq a_i \leq \overline{\kappa} \label{eq:constr_level}\\
    & \underline{\kappa} \leq \kappa(\lambda_i;\mathbf{\Theta}^0) \leq \overline{\kappa} \label{eq:constr_curv}
\end{align}
\end{subequations}
for $i,j \in \mathbb{I}_{[1,n_0]}$. \textcolor{black}{These constraints ensure that each sigmoid represents a single curvature transition with physically plausible parameters, by enforcing that the transition points are ordered, between minimal and maximal arc length (constraint \eqref{eq:constr_ordering}), and that the curvature change and absolute curvature remain within valid bounds (constraints \eqref{eq:constr_level} and \eqref{eq:constr_curv}).} Additional polytopic constraints on $c_i$ can also be incorporated as needed, as well as any prior information constraining the Cartesian coordinates of the map points, which can be imposed by $\mathcal{P}$. \textcolor{black}{We note that by additionally constraining the curvature of the reference trajectory to not exceed $\frac{1}{\eta_{max}}$, where $\eta_{max}$ is the maximal allowed lateral distance of the car, the condition for unique  Frenet frame representation is satisfied~\cite{wursching2024robust}. }

\subsection{\textcolor{black}{Receding-Horizon Map Update}}
\label{subsec:est}

{\color{black} At each time step, the road model is updated using newly acquired centerline measurements while maintaining a finite representation of the road ahead of the vehicle. Rather than constructing a global map, the proposed method operates on a receding spatial horizon, resulting in two advantages. First, it prevents the number of model parameters from growing unbounded as the vehicle progresses. Second, it avoids the accumulation of long-term estimation errors, as the map is continuously updated with current perception data. As a result, drift effects associated with global map building are avoided.

Let $\mathbf{\Theta}^k$ denote the current road model and $\mathbf{P}^{k+1}$ the newly observed centerline points. The model is updated only if the new measurement provides sufficient and consistent information. To this end, we evaluate the overlap between $\mathbf{P}^{k+1}$ and the current map $\mathbf{m}^k$,
where a measurement point is considered overlapping if its Euclidean
distance to any map point is below a defined threshold. Two conditions must be satisfied: (i) the number of overlapping points must exceed a predefined threshold, ensuring consistency with the existing estimate, and (ii) the number of non-overlapping (novel) points must also exceed a threshold, ensuring that new information is incorporated. If either condition is not met, the update is skipped and $\mathbf{\Theta}^{k+1} = \mathbf{\Theta}^k$, preventing unnecessary computation and
avoiding degradation of the map with unreliable measurements. If both conditions are satisfied, the update proceeds, and the optimization is performed over a combination of retained map points and newly observed points. Specifically, we retain $\bar{L}_k$ points from the previous map $\mathbf{m}^k$ and augment them with $\bar{M}_{k+1}$ non-overlapping points from $\mathbf{P}^{k+1}$. The retained map segment is selected to include only the portion of the road ahead of the vehicle, while map points corresponding to curvature transitions behind the current vehicle position are discarded, along with their associated sigmoid parameters. This procedure ensures that the number of active basis functions remains bounded over time. If no
successful update occurs for a prolonged period, the map can be discarded
entirely and the estimation reinitialized from the current observation
using the procedure in Section~\ref{subsec:init_est}.

Given  $\bar{L}_k$ retained map points from $\mathbf{m}^k$ and the $\bar{M}_{k+1}$ new, non-overlapping measurement points from $\mathbf{P}^{k+1}$, the updated curvature parameters $\mathbf{\Theta}^{k+1}$ are obtained by solving
}

\begin{equation}
\begin{aligned}
    \min_{\mathbf{Z}^{k+1}}&\sum_{i = 0}^{\bar{L}_k{-}1}\lVert {{m}_{i}^{k}{-}{{m}_{i}^{k{+}1}} \lVert}^2{+}\!\!\!\!\sum_{i = 0}^{\bar{M}_{k+1}{-}1}\lVert {{p}_{i}^{k{+}1}{-}{{m}_{i{+}\bar{L}_k}^{k{+}1}} \lVert}^2{+}\mathcal{L}(\mathbf{\Theta}^{k{+}1})\\
\text{s.t. } & {m_0^{k+1}}={m_0^{k}} \\
&  \textcolor{black}{\lambda_0 =s_k} \\
& \lambda_{i+1} = \lambda_i + \Delta \lambda_i \\ 
&  \begin{bmatrix}
           \alpha_{i+1}^{k+1} \\
           {m_{i+1}^{k+1}}
         \end{bmatrix} = f_{\Delta}\left(\begin{bmatrix}
           \alpha_{i}^{k+1} \\
           {m_{i}^{k+1}}
         \end{bmatrix}, \kappa( \lambda_i; \mathbf{\Theta}^{k+1}) \right)\\
&\textcolor{black}{\alpha_0^{k+1} \in (-\pi, \pi], {m_i^{k+1}} \in \mathcal{P}, \mathbf{\Theta}^{k+1} \in \mathcal{Q}}, 
\end{aligned}
\label{eq:map_update}
\end{equation}
where $\mathbf{Z}^{k+1} = [\alpha_0^{k+1}, {\mathbf{\Theta}^{k+1}}^\top, {\mathbf{m}^{k+1}}^\top]^\top$, $\alpha_0^{k+1}~\in~\mathbb{R},$ $\mathbf{\Theta}^{k{+}1}~\in~\mathbb{R}^{3n_{k+1}+1}, \mathbf{m}^{k+1} \in \mathbb{R}^{2 \times (\bar{L}_k + \bar{M}_{k+1}))}$.  Similarly to the previous subsection, parameter $\Delta \lambda_i$ denotes arc length between ${m_{i}^k}$ and ${m_{i+1}^k}$ for $i \leq \bar{L}_k-1$, and ${p_{i-\bar{L}_k}^{k+1}}$ and ${p_{i+1-\bar{L}_k}^{k+1}}$ for $i \geq \bar{L}_k$. In both cases, the arc length between the points is approximated by their Euclidean distance.  \textcolor{black}{The first term in the cost enforces consistency with the retained portion of the map, while the second term aligns the model with newly observed measurements. The constraints are the same as in~\eqref{eq:constrs}.}

\begin{algorithm}[tbp]
    \caption{Curvature Estimation Algorithm at time $k$}\label{al:1}
    \begin{algorithmic}[1]
    \State \textbf{Input at time $k$:}  $x^{c}(k), \mathbf{P}^k, \mathbf{\Theta}^{k-1}, \mathbf{\mu}^{k-1}$
    \State \textbf{Output:} $\mathbf{\Theta}^{k}, \mathbf{\mu}^{k} $
    \If{$k==0$} 
    \State $\mathbf{\Theta}^{0}, \mathbf{\mu}^{0}  \leftarrow$ solve \eqref{eq:map_init} using $\mathbf{P}^0$
    \Else    
        \State $ r, overlap \leftarrow\text{DetermineOverlap}(\mathbf{P}^k, \mathbf{\Theta}^{k-1}, \mathbf{\mu}^{k-1}) $
        \If{ $overlap$}
            \State $\overline{\mathbf{P}}^k, \overline{\mathbf{\Theta}}^{k-1}, \overline{\mathbf{\mu}}^{k-1} \leftarrow\text{RM}(x^{c}(k), \mathbf{P}^k, \mathbf{\Theta}^{k-1}, \mathbf{\mu}^{k-1}, r) $
            \textcolor{black}{\State  $\mathbf{\Theta}^{k}_{init}  \leftarrow\text{InitTheta}(\overline{\mathbf{P}}^k, \overline{\mathbf{\mu}}^{k-1})$}
            \State  $\mathbf{\Theta}^{k}, \mathbf{\mu}^{k}  \leftarrow$ solve \eqref{eq:map_update} using $\overline{\mathbf{P}}^k, \overline{\mathbf{\Theta}}^{k-1}$, $\overline{\mathbf{\mu}}^{k-1}$, \textcolor{black}{$\mathbf{\Theta}^{k}_{init}$}
        \Else
         \State $\mathbf{\Theta}^{k}, \mathbf{\mu}^{k} \leftarrow \mathbf{\Theta}^{k-1}, \mathbf{\mu}^{k-1}$
        \EndIf
    \EndIf
    \end{algorithmic}
\end{algorithm}

\subsection{Curvature Estimation Summary}
\label{subsec:algo}

{\color{black} Algorithm~\ref{al:1} summarizes the complete curvature estimation procedure at each time step~$k$. The inputs consist of the current Cartesian vehicle state $x^c(k)$, the newly observed centerline points $\mathbf{P}^k$, and the previously estimated model $\mathbf{\Theta}^{k-1}$ with its corresponding reconstruction $\mathbf{\mu}^{k-1}$. At $k=0$, the curvature parameters are obtained by solving the initial curvature estimation problem~\eqref{eq:map_init}. For $k > 0$, the algorithm evaluates the overlap between the new measurement $\mathbf{P}^k$ and the existing map using  \texttt{DetermineOverlap} to determine whether the observation provides sufficient novel information for a model update. If so, the receding-horizon retain-map procedure (\texttt{RM}) (i) extracts the non-overlapping (novel) measurement points $\overline{\mathbf{P}}^k$ from ${\mathbf{P}}^k$, and (ii) retains the relevant portion of the previous map $\overline{\mathbf{\mu}}^{k-1}$ together with the associated curvature parameters $\overline{\mathbf{\Theta}}^{k-1}$. Then, an optional initialization routine (\texttt{InitTheta}) provides a warm start, and the updated parameters $\mathbf{\Theta}_k$ are obtained by solving~\eqref{eq:map_update}. If the overlap conditions are not met, the update is skipped and $\mathbf{\Theta}^k = \mathbf{\Theta}^{k-1}$, preventing the incorporation of unreliable or uninformative measurements. The output is the curvature model $\mathbf{\Theta}_k$ and the associated centerline reconstruction $\boldsymbol{\mu}_k$, which serve as input to the uncertainty sampling module described next.
}

\section{Uncertainty Sampling Based on Hausdorff Distance}
\label{sec:uncert_quant}

{\color{black}

As motivated in Section~\ref{sec:overview}, we adopt a formulation in which perception uncertainty is represented at the geometric level. Let $\mathbf{m}^k$ denote the nominal centerline obtained from the estimated curvature parameters $\mathbf{\Theta}^k$. We assume that the true centerline lies within a bounded neighborhood of $\mathbf{m}^k$, defined via some similarity metric $d_S(\cdot)$. This results in a geometric ambiguity set
\[
\mathcal{M}^k = \{ \tilde{\mathbf{m}} \;|\; d_S(\mathbf{m}^k, \tilde{\mathbf{m}}) \leq D^k \},
\]
where $D^k > 0$ denotes the expected geometric estimation error.

The parameter $D^k$ can be interpreted as the radius of a tube around the nominal centerline estimate within which the true road is assumed to lie. It represents a cumulative bound capturing perception noise, lane detection errors, and road model mismatch. \textcolor{black}{In practice, $D^k$ is calibrated from data by running the 
curvature estimation module on representative trajectories 
and recording the worst-case similarity metric between 
estimated and ground-truth centerlines. If certified error 
bounds from the perception module are available, they can 
be used directly.}

To approximate the set $\mathcal{M}^k$, we generate multiple centerline realizations by perturbing the curvature parameters and retaining only those that satisfy the similarity constraint. The procedure is summarized in Algorithm~\ref{al:2}.

At each iteration, a candidate parameter vector $\overline{\mathbf{\Theta}}^k_i$ is generated by adding zero-mean Gaussian noise with covariance $\Sigma_{\mathbf{\Theta}^k}$ to the nominal parameters $\mathbf{\Theta}^k$. The corresponding centerline $\overline{\mathbf{m}}$ is then reconstructed using the discrete propagation model~\eqref{eq:forw_prop1}, initialized with $m_0^k$ and $\alpha_0^k$ extracted from $\mathbf{\mu}^k$. The candidate is accepted only if it satisfies the similarity constraint $d_S(\mathbf{m}^k, \overline{\mathbf{m}}) \leq D^k$, ensuring that all retained samples lie within the prescribed geometric uncertainty set. This procedure is repeated $N_{\text{rep}}$ times, resulting in a set of admissible candidates. From this set, $N_s$ representative realizations are selected using a selection strategy (function \texttt{Select}), e.g., by retaining the candidates with 
the largest Hausdorff distance from the nominal centerline 
to prioritize geometrically challenging scenarios.

During the sampling, all curvature parameters within the current spatial horizon are perturbed. This avoids imposing structural assumptions about where uncertainty arises and ensures that the sampling captures the full geometric variability of the estimated road. The dimensionality of the sampling space remains bounded due to the receding-horizon formulation of the map.
}

{\color{black} We adopt the Hausdorff distance as the similarity metric $d_{S}(\cdot)$, as it captures the worst-case pointwise deviation between centerlines and is therefore well suited for safety-critical constraint enforcement. For two non-empty subsets $A, B \subseteq {X}$ of a metric space $X$ with a distance function $d: X \times X \to \mathbb{R}$, it is defined as
\begin{equation}
    d_H(A, B) = \max \left\{ \sup_{a \in A} \inf_{b \in B} d(a,b),\; \sup_{b \in B} \inf_{a \in A} d(a,b) \right\}.
\end{equation}}

\vspace{-0.3cm}
{\color{black}
The proposed approach constitutes a set-based uncertainty model rather than a probabilistic one. The covariance $\Sigma_{\mathbf{\Theta}^k}$ does not encode a statistical model of perception noise, but serves as a structured mechanism to explore local variations in the parameter space. The practical robustness of the method is thus determined only by the induced geometric deviation, controlled by $D^k$. Furthermore, the number of sampling attempts $N_{\text{rep}}$ and the number of retained realizations $N_s$ govern the trade-off between computational complexity and coverage of the uncertainty set. Increasing $N_{\text{rep}}$ improves the likelihood of obtaining diverse admissible samples, while $N_s$ determines how many representative scenarios are passed to the controller.

Although $D^k$ and $\Sigma_{\mathbf{\Theta}^k}$ serve distinct 
purposes, i.e., the former defines the admissible set and governs controller 
conservatism, while the latter controls the versatility of samples, they are operationally 
coupled. The covariance should be scaled in proportion to the Hausdorff bound: if 
$D^k$ is large, a larger $\Sigma_{\mathbf{\Theta}^k}$ is needed so 
that the sampled realizations cover the larger extent of the uncertainty tube; if $D^k$ 
is small, a smaller $\Sigma_{\mathbf{\Theta}^k}$ avoids excessive rejections and 
maintains sampling efficiency. In the current implementation, both parameters are 
fixed offline, during the calibration procedure. Adapting $D^k$ online based on a perception confidence indicator derived from the mismatch of the fitted centerline and the measurements, and 
co-adapting $\Sigma_{\mathbf{\Theta}^k}$ to preserve coverage and acceptance rate, 
constitutes a natural extension of the framework.

The proposed method therefore provides a tractable mechanism for propagating perception uncertainty by constructing a finite set of geometrically consistent centerlines. The resulting samples are used in the controller to ensure safe operation across multiple plausible interpretations of the environment. }\textcolor{black}{We note that the proposed sampling scheme is a practical approach inspired by the scenario framework, and that establishing formal theoretical connections remains \textcolor{black}{a relevant} direction for future work.}

\begin{algorithm}[tbp]
    \caption{Sampling of Centerline Realizations}\label{al:2}
    \begin{algorithmic}[1]
    \State \textbf{Input at time $k$:} {$\mathbf{\Theta}^{k},  \mathbf{\mu}^{k}, \Sigma_{\mathbf{\Theta}^k}, D^k, N_{\text{rep}}, N_s$}
    \State \textbf{Output:} $[\overline{\mathbf{\Theta}}^{k}_1, \cdots, \overline{\mathbf{\Theta}}^{k}_{N_s}] $
    \State $i \leftarrow 0$
    \State {$CandidateList \leftarrow []$}
   \While{$i<N_{\text{rep}}$}
            \State {$\overline{\mathbf{\Theta}}^{k} \leftarrow \mathbf{\Theta}^{k} + \text{GaussianNoise}(\Sigma_{\mathbf{\Theta}^k})$}
            \State $\overline{\mathbf{m}} \leftarrow$  \textcolor{black}{obtain Cartesian coordinates from curvature} 
            \State \textcolor{black}{parameters using discretized} $\eqref{eq:forw_prop1}$ and $m_0^k, \alpha_0^k,  \overline{\mathbf{\Theta}}^{k}$
            \If {${d_S}(\mathbf{m}^k, {\overline{\mathbf{m}}})\leq D^k$} 
                \State insert $\overline{\mathbf{\Theta}}^k$ into {$CandidateList$}
            \EndIf
            \State $i\leftarrow i+1$
    \EndWhile
    \State {$[\overline{\mathbf{\Theta}}^{k}_1, \cdots, \overline{\mathbf{\Theta}}^{k}_{N_s}] \leftarrow \text{Select}(CandidateList,N_s)$} 
    \end{algorithmic}
\end{algorithm}

\section{\textcolor{black}{Uncertainty-Aware Control}}
\label{sec:rob_ctrl}

{\color{black} 

Building on the bounded centerline uncertainty set constructed
in Section~\ref{sec:uncert_quant}, we now incorporate these
curvature realizations into the control formulation. We note that the control formulation is agnostic to the specific method
used to generate these curvature realizations: it requires only a finite
set of bounded and differentiable curvature profiles as input, and is
therefore compatible with any upstream estimation or uncertainty
quantification approach that produces such a set. At time
step $k$, the Cartesian vehicle state $x^c(k)$ is projected onto
each of the $N_s$ sampled centerlines, corresponding to curvature
parameter realizations
$\overline{\mathbf{\Theta}}^k_1, \dots, \overline{\mathbf{\Theta}}^k_{N_s}$.
This yields $N_s$ initial states in the Frenet frame, denoted by
$x^1(k), \dots, x^{N_s}(k)$, each associated with a different
plausible centerline. This coordinate transformation is performed
once per time step and the resulting states are used throughout
the optimization, i.e., no repeated Cartesian-to-Frenet
conversions are required within the solver loop. 

To impose constraint satisfaction across all centerline
realizations, we formulate a scenario-based extension of the
MPCC introduced in Section~\ref{subsec:MPCCF}. A single
control sequence $\mathbf{U}$ is applied to $N_s$ parallel system
realizations, each governed by the dynamics $f^d(\cdot)$ with a
different curvature profile
$\kappa(\cdot;\, \overline{\mathbf{\Theta}}^k_l)$, producing $N_s$
predicted trajectories
$\overline{\mathbf{X}} = [{\overline{\mathbf{X}}}^{\top}_1,
\cdots, {\overline{\mathbf{X}}}^{\top}_{N_s}]^{\top}$. \textcolor{black}{We obtain the control input $\mathbf{U}$ by solving the following
optimization problem in receding horizon fashion}

\begin{equation}
\begin{aligned}
    \min_{\overline{\mathbf{X}}, \mathbf{U}} & \sum_{i = 0}^{N-1} \overline{l}_i(\overline{\mathbf{x}}_i, u_i) + \overline{L}_N(\overline{\mathbf{x}}_N)
\\
\text{s.t. } 
& \overline{x}^l_0 = {x}^l(k), \quad l \in \mathbb{I}_{[1,\textcolor{black}{N_s}]}\\
& \overline{x}^l_{i+1} = f^d(\overline{x}^l_i, u_i, \kappa_l(\cdot)), \quad l \in \mathbb{I}_{[1,\textcolor{black}{N_s}]} \\ 
& \overline{x}^l_i \in \overline{\mathcal{X}}, \quad l \in \mathbb{I}_{[1,\textcolor{black}{N_s}]}\\
& \overline{x}^l_N \in \overline{\mathcal{X}}_f, \quad l \in \mathbb{I}_{[1,\textcolor{black}{N_s}]}\\
& u_i \in \mathcal{U},
\end{aligned}
\label{eq:pa-control}
\end{equation}
where, to simplify notation, we denote ${\kappa}_l(\cdot)={\kappa}({ \cdot; \overline{\mathbf{\Theta}}}^k_l), \overline{\mathbf{X}}_l = [{\overline{x}_0^l}^{\top}, \cdots, {\overline{x}_N^l}^{\top}]^{\top}$ for $l \in \mathbb{I}_{[1,N_s]}$ and $\overline{\mathbf{x}}_i=[{\overline{x}_i^1}^{\top}, \cdots, {\overline{x}_i^{N_s}}^{\top}]^{\top}$. The cost is chosen as the average over all trajectories, i.e., $\overline{l}_i(\overline{\mathbf{x}}_i, {u}_i)= \frac{1}{N_s} \sum_{l=1}^{N_s} l_i(\overline{x}_i^l, u_i)$ and $\overline{L}_N(\overline{\mathbf{x}}_N)~=~\frac{1}{N_s} \sum_{l=1}^{N_s} L_N(\overline{x}_N^l)$. \textcolor{black}{
While the constraints from Section~\ref{subsec:MPCCF} 
remain, additional ones are imposed to ensure the validity 
of the vehicle's arc length coordinate along the prediction 
horizon, and prevent the car from reaching poses beyond the currently perceived portion of the road. Specifically, the state and terminal 
constraint sets are augmented as 
$\overline{\mathcal{X}} = \mathcal{X} \cap \{x \mid 0 \leq s 
\leq s_{\max}\}$ and 
$\overline{\mathcal{X}}_f = \overline{\mathcal{X}}$, where 
$s$ denotes the vehicle's progress along the reference path 
(first component of the Frenet-frame state) and $s_{\max}$ 
is the maximum arc length of the perceived centerline at 
the current time step. } 

{\color{black}

The scenario-based formulation exploits the structure of the Frenet-frame dynamics to limit computational growth. The vehicle dynamics admit an eight-dimensional state $x = (s, \eta, \phi, v_x, v_y, r, \delta, \tau)$ and a two-dimensional input $u = (\Delta\delta, \Delta\tau)$. Of the eight state dimensions, only the first three depend on the curvature of the reference path (cf.~\eqref{eq:curv_dyn}) and therefore differ across scenarios. The remaining five states are shared across all realizations. The augmented state dimension per stage is thus $3 N_s + 5$ rather than $8 N_s$, and the number of dynamics constraints and state constraints on $(s, \eta, \phi)$ scales as $N_s \cdot N$, while input constraints and shared-state constraints remain independent of $N_s$.

Each sampled centerline corresponds to a plausible realization of the road geometry and induces a different vehicle trajectory. By enforcing constraints across all such realizations, the controller explicitly accounts for perception uncertainty. The parameter $D^k$, introduced in Section~\ref{sec:uncert_quant}, directly influences the controller behavior. Larger values of $D^k$ enlarge the set of admissible centerlines and result in more conservative control actions, while smaller values reduce conservatism when perception is more reliable. In this sense, the approach provides a systematic mechanism to trade off performance and practical robustness based on estimated perception accuracy.

}

\section{Numerical Results}
\label{sec:Results}

{\color{black} In the following, we present the results for each stage of the proposed method. Section~\ref{subsec:sim_setup} introduces the simulation setup, including two tracks with piecewise-constant curvature used throughout the evaluation. Section~\ref{subsec:est_res} evaluates the curvature estimation module in this simulated setting, reporting estimation errors with respect to the ground-truth. To further validate the robustness of this module to realistic visual noise, depth inaccuracies, and imperfect centerline detection, we additionally evaluate it on real-world images collected from an onboard camera on an F1TENTH vehicle navigating a physical track. Section~\ref{subsec:sampling_res} demonstrates that the proposed sampling procedure generates geometrically consistent centerline realizations forming a bounded set around the nominal estimate. Finally, Section~\ref{sec:control_res} evaluates the uncertainty-aware controller in simulation and shows that incorporating multiple centerline realizations into the MPC formulation reduces lateral deviation from the true road centerline compared to the nominal controller.}

\subsection{Simulation Setup}
\label{subsec:sim_setup}

{\color{black}\textcolor{black}{We evaluate our approach on two tracks with piecewise-constant curvature (see Figure~\ref{fig:tracks}) and constant,
known width. The tracks are synthetic, rendered in Blender~\cite{blender}, with a clearly visible centerline,
no lane boundary markings, no clutter, and no other dynamic
agents. The vehicle's Cartesian state is assumed known at every
time step. These simplifications are adopted to isolate the contribution of perception-induced geometric uncertainty to closed-loop control performance, framing the present study as a methodological validation of the proposed pipeline; relaxations of each assumption are discussed in Section~\ref{sec:Disc}. Despite these simplifications, the tracks were selected to be
challenging from a control perspective, featuring sharp turns and frequent curvature changes typical of realistic miniature car racing scenarios~\cite{carron2023chronos}. Incorporating perception further increases the difficulty, as the rapid curvature variations, combined with the low mounting position of the onboard camera, make centerline estimation more error-prone. Additionally, because only partial track information is available ahead of the vehicle at any given time, the system must often reduce speed to avoid unsafe behavior.}

{In this simulated setting, the curvature estimation module is subject to errors arising from imperfect centerline pixel extraction and coordinate quantization inherent to the detection method. The cumulative effect of these errors together with realistic image noise, depth inaccuracies, and visual artifacts is evaluated separately on real-world data in Section~\ref{subsec:est_res}.}

Two examples of the resulting visual measurements are 
shown in Figure~\ref{fig:Simulation}, showcasing different 
views of the white centerline of $\SI{2}{cm}$ width. } The camera's axis is kept parallel to the floor and aligned with the car's orientation, with a perspective lens set to a focal length of $\SI{15}{mm}$. At each time step, an RGB-D image with resolution $1280 \times 720$ pixels is rendered using Blender.

\begin{figure}
    \centering
    \includegraphics[width=0.99\linewidth]{ 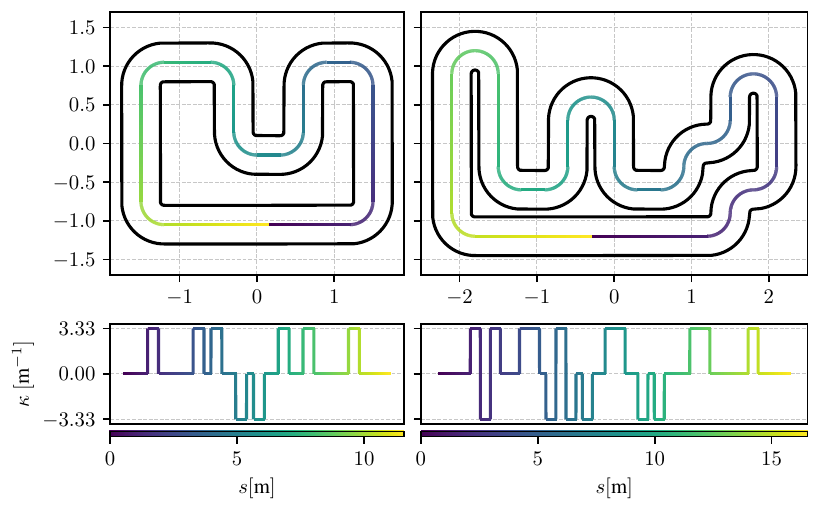}
    \caption{Top: Track A (left) and track B (right). Bottom: curvature of each track as a function of arc length.}
    \label{fig:tracks}
\end{figure}

The car dynamics are simulated in Cartesian coordinates as in \cite{tearle2021predictive}. The simulation uses the following car parameters corresponding to the miniature racing cars from \cite{carron2023chronos}: $m=\SI{0.2}{kg}, I_z~=~\SI{0.0004}{kg m^2}, l_r~=~\SI{4.5}{cm}, l_f~=~\SI{5.6}{cm}$, $B_r~=~8, B_f=8, C_r=1.7, C_f=1.4, D_r=\SI{0.6}{N}, D_f=\SI{0.43}{N}$, $C_1 = \SI{0.98}{N}, C_2 = C_3 = 0, C_4 = \SI{0.03}{kg \cdot m}, C_5= \SI{0.02}{\frac{kg}{s}}$ and $C_6= \SI{0.08}{N}$. The sampling frequency is set to $f_s=\SI{30}{Hz}$. 

At each time step, given the car's current state and its centerline estimate, an open-loop input sequence is computed over a prediction horizon of $N=35$ using the bicycle model of the car in Frenet frame, and applied to the car in a receding horizon fashion. We consider two different controllers, a  \textit{nominal} and an \textit{uncertainty-aware} one. The former is the curvilinear MPCC controller introduced in Section \ref{subsec:MPCCF}. \textcolor{black}{It is configured with a road width of $W=\SI{0.5}{m}$, and cost weights $q_s = 100, q_{\eta}=75, q_{\phi}=1000$, $R=\text{diag}(0.01, 0.001)$, and $q_{vx}=q_{vy}=10$. State and input constraints are imposed by setting $  \SI{0.2}{\frac{m}{s}} \leq v_{x,i} \leq \SI{5}{\frac{m}{s}}$, $ \SI{-1}{\frac{m}{s}} \leq v_{y,i} \leq  \SI{1}{\frac{m}{s}}$, $  \SI{-5}{\frac{rad}{s}} \leq r_i \leq  \SI{5}{\frac{rad}{s}}$, $  \SI{-0.41}{rad} \leq \delta_i \leq  \SI{0.41}{rad}$, $  0 \leq \tau_i \leq 0.5$ for $i \in \mathbb{I}_{[0,N]}$ and $  \SI{-5}{rad} \leq \Delta \delta_j \leq  \SI{5}{rad}$, $  -5 \leq \Delta \tau_j \leq  5$, where input to the system at time $j \in \mathbb{I}_{[0, N-1]}$ along the horizon is given by $[\Delta \delta_j, \Delta \tau_j]^{\top}$.} The \textit{uncertainty-aware} controller is described in Section \ref{sec:rob_ctrl} and employs the same cost weights and terminal constraints as the nominal one, with a few exceptions detailed in Section~\ref{sec:control_res}. Both controllers use IPOPT \cite{wachter2006implementation} as  solver.

\begin {figure}[t]
\centering
\vspace{-0.3cm}
\begin{adjustbox}{max height=0.55\textwidth, max width=0.48\textwidth}
\begin{tikzpicture}[scale=1.0]

    \node (pic1) at (-2.2, 0.0) {\includegraphics[height=.17\textwidth,width=.23\textwidth]{ 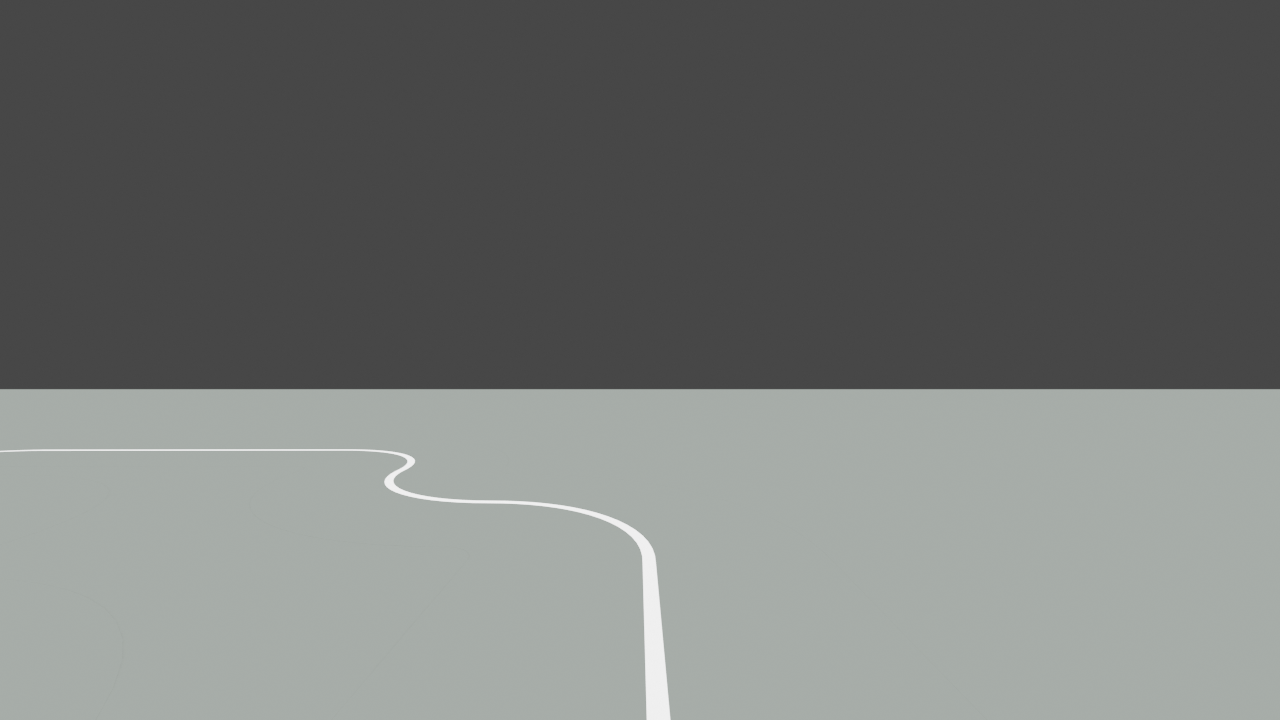}};

    \node (pic2) at (2.2, 0.0) {\includegraphics[height=.17\textwidth,width=.23\textwidth]{ 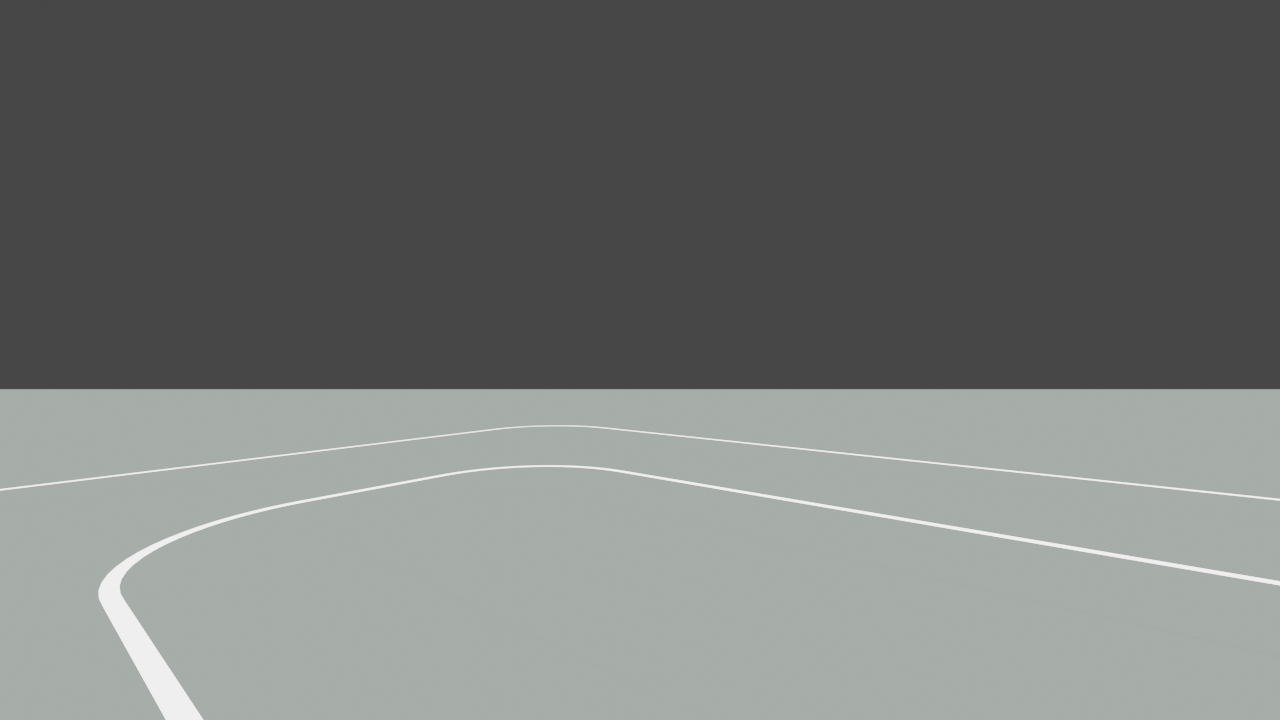}};

\end{tikzpicture}
\end{adjustbox}
    \caption{Different views acquired along track B.} 
\label{fig:Simulation}
\vspace{-0.5cm}
\end{figure}

\subsection{Centerline Estimation}
\label{subsec:est_res}

{\color{black} In this subsection, we present the results of the proposed centerline estimation method. We first evaluate it in the simulated setting described in Section~\ref{subsec:sim_setup}, then validate it on real-world images collected from an F1TENTH vehicle. In both cases, centerline point clouds are extracted from RGB-D measurements using edge drawing~\cite{topal2012edge}, which has been demonstrated to generalize to realistic road imagery~\cite{8938821}. The extracted point clouds are then processed by the curvature estimation procedure outlined in Algorithm~\ref{al:1}.

}

\subsubsection{\textcolor{black}{Simulation Results}}
\label{subsec:curv_est_sim}

To {initially} decouple perception from control, we employ a nominal controller that utilizes ground-truth track information to guide the car. This ensures that the car follows a realistic trajectory independent of perception errors. However, to maintain a realistic perception-based scenario at each time step, the controller only has access to ground-truth curvature up to the point corresponding to the highest arc length of the estimated centerline.

{In the curvature estimation method, we set  $ \overline{\kappa}=\SI{4}{m^{-1}}$ and $\underline{\kappa}=\SI{-4}{m^{-1}}$.} This choice is guided by the condition from \cite{wursching2024robust}, where it is shown that unique representation with respect to the curve can be achieved for a point with curvilinear coordinates $s$ and $\eta$ only if it holds that $\eta \kappa(s, \mathbf{\Theta}) \leq 1$. Since in our case $\eta \leq \frac{W}{2} = \SI{25}{cm}$, it has to hold that $\overline{\kappa} \leq \SI{4}{m^{-1}}$ and similarly, $\underline{\kappa} \geq \SI{-4}{m^{-1}}$, resulting in the bounds that contain the maximum track curvature of $\SI{\pm 3.33}{m^{-1}}$ (see Figure \ref{fig:tracks}). Furthermore, the overlap of the old map and new measurement is considered as sufficient if it contains at least 30 of the observation points, and the update is performed if at least 30 points of the observation are not a part of this overlap. Each map is limited in length by $\SI{3.5}{m}$, i.e., all observation points beyond this map length are ignored. \textcolor{black}{The specific map length was chosen to balance two competing objectives: longer maps enlarge the optimization problem and increase computational cost, while shorter maps may restrict the feasible motion of the vehicle and limit MPC performance. } Furthermore, as the road is piecewise-constant in curvature, we set all sigmoid steepness parameters $c_i$, which describe the steepness of the curvature change, to $30$ and optimize for $\kappa_0$ and all $a_i$ and $b_i$.

To reduce computational complexity, we perform subsampling of each point cloud $\mathbf{P}^k$ such that the distance between consecutive points is as close to $\Delta \lambda = \SI{4}{cm}$ as possible, and use only these subsampled points in the road model optimization process. \textcolor{black}{To initialize problems~\eqref{eq:map_init} and \eqref{eq:map_update}, parameters corresponding to the retained map segment are assigned their previously estimated values. For newly observed points, 
curvature is computed directly from the Cartesian coordinates 
using 
\begin{equation}
\kappa=\frac{x^{\prime} y^{\prime \prime}-y^{\prime} x^{\prime \prime}}
{\left(x^{\prime 2}+y^{\prime 2}\right)^{3 / 2}},
\label{eq:curvature_form}
\end{equation}
where the derivatives are approximated via finite differences.  The number and approximate placement of sigmoids are determined by thresholding the resulting curvature values to detect transitions. Although this direct computation is noise-sensitive, it suffices for determining an approximate sigmoid configuration, which provides an effective starting point for the solver. For segments where no curvature change is detected, a single sigmoid with zero amplitude is used.  Each optimization problem is solved using IPOPT \cite{wachter2006implementation} and the same parameter settings were used across both simulated tracks.} 

\begin {figure}[t]
\centering
\begin{adjustbox}{max height=0.55\textwidth, max width=0.48\textwidth}
\begin{tikzpicture}[scale=1.0]

    \node (pic1) at (-2.2, 0.0) {\includegraphics[height=.17\textwidth,width=.23\textwidth]{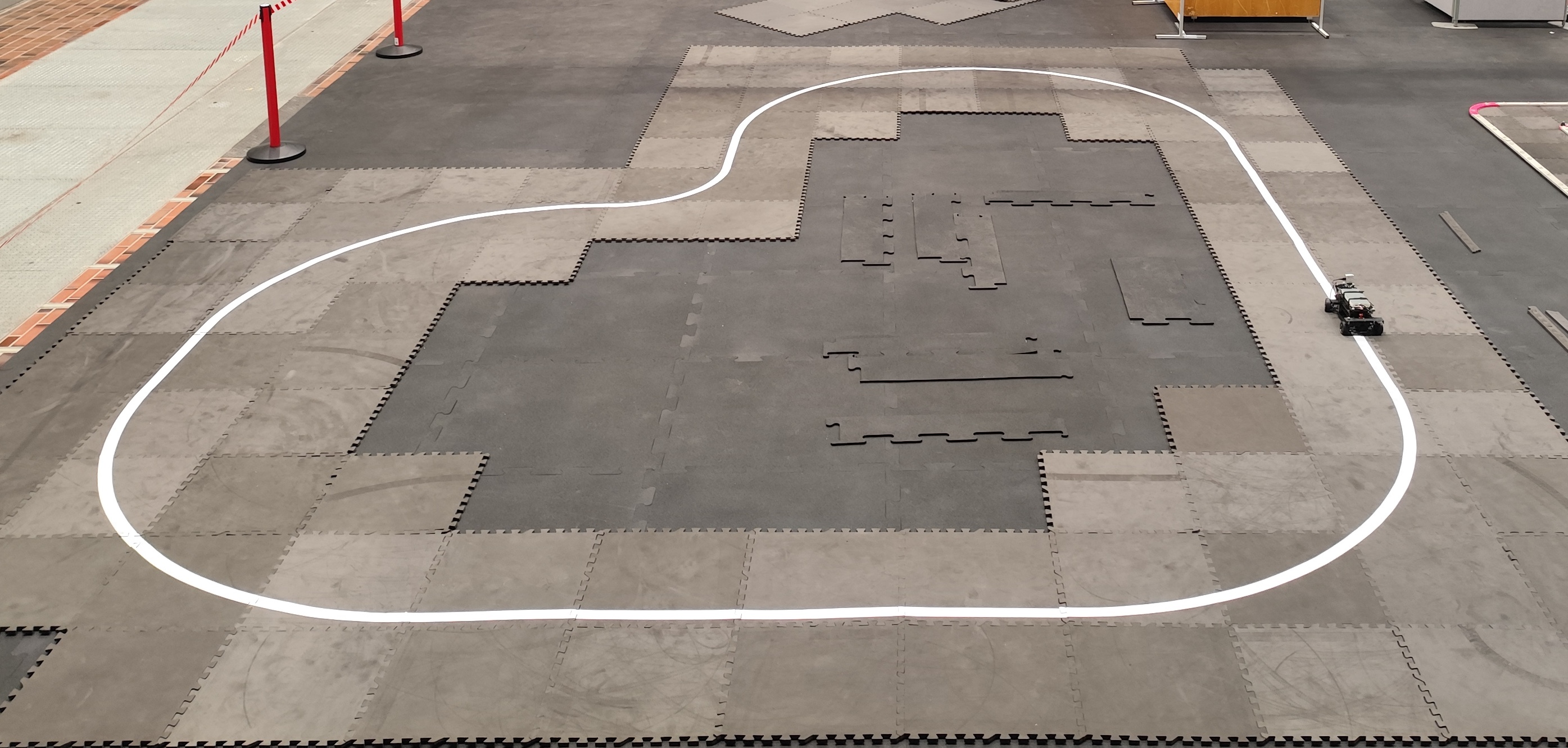}};

    \node (pic2) at (2.2, 0.0) {\includegraphics[height=.17\textwidth,width=.23\textwidth]{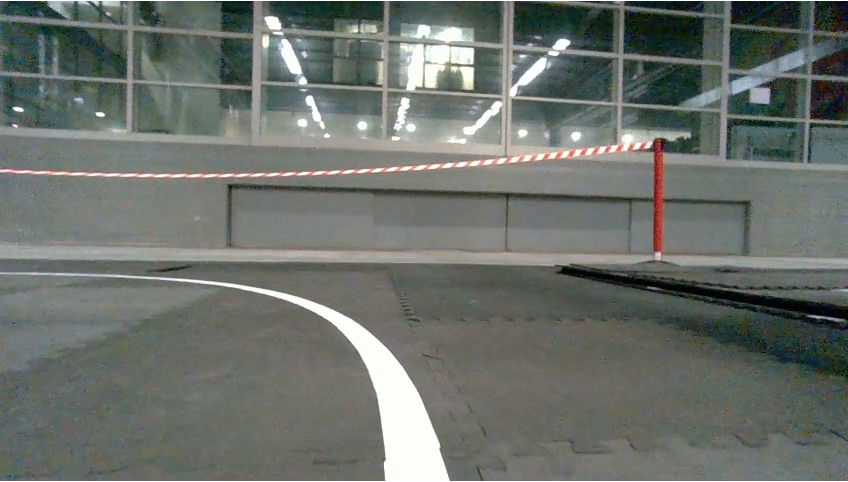}};

\end{tikzpicture}
\end{adjustbox}
    \caption{\textcolor{black}{The real-world track (left) and the view along it (right).}} 
\label{fig:phys_track}
\end{figure}

{We evaluate our curvature estimation method in an ablation study.} To ensure different viewpoints between multiple runs around the track, we evaluate the proposed method and the baselines for 10 different trajectories on each track, achieved by perturbing the initial state of the car, and the parameters of the controller (the cost weights, as well as the mass and the yaw moment of inertia of the car). {Table \ref{tab:ablation} presents the results of the ablation study, including the mean Hausdorff distance between the estimated centerline and the ground-truth across all centerline estimates, as well as the mean absolute curvature error averaged over all trajectories. We also report the success rate (SR column) of each method, indicating how many out of the 10 trajectories resulted in a complete track estimation.} 

\begin{table*}[t!]
    \caption{\textbf{\textsc{Ablation study}} \textcolor{black}{We run the baselines (naive and smooth naive) and the proposed method with and without initialization for 10 different trajectories on each of the tracks.}  We report average and standard deviation of the average Hausdorff distance for each trajectory. We also compute the mean and standard deviation of the mean absolute error in curvature over all trajectories ($\kappa$ MAE). In the last column, we provide the worst case Hausdorff distance over all estimated centerlines for all trajectories on both tracks. 
    }
    \label{tab:object_full_evol}
        \setlength{\tabcolsep}{3pt}
        \centering
        \begin{tabular}{|l|l|l|c|c|c||c|c|c||c|}
        \toprule
      &  & & \multicolumn{2}{c||}{Track A} & \multicolumn{3}{c||}{Track B} & \multicolumn{1}{c|}{Worst case}\\ \midrule
    {Method}  & {Init} & HD  $[cm]$ $\downarrow$ & $\kappa$ MAE [$m^{-1}$] $\downarrow$ & SR $[\%]$ $\uparrow$ &   HD  $[cm]$ $\downarrow$ & $\kappa$ MAE [$m^{-1}$] $\downarrow$ & SR $[\%]$ $\uparrow$ &   HD  $[cm]$ $\downarrow$ \\
 \midrule
        Naive & -   & 3.65  {$\pm 0.50$}&  $2.36$  {$\pm 0.16$} & $\mathbf{100}$ & 4.87  {$\pm 0.48$} & $3.25$  {$\pm 0.10$} & 90 & \textbf{8.06}  \\
        Smooth naive & -    &{\textbf{3.27}  {$\mathbf{\pm 0.50}$}}&   {$1.78$  {$\pm 0.07$}}  &  $\mathbf{100}$ &{9.02  {$\pm 2.53$}}&   2.46  {$\pm 0.11$}& 80 & $17.40$ \\
         \midrule
        Ours & \xmark  &  6.06  {$\pm 2.45$}&  $1.03$  {$\pm 0.25$} & 90 & 6.79  {$ \pm 2.05$}&   1.19  {$\pm 0.24$} & 50 & 28.70  \\
        % Ours & \xmark   & 7.99  {$\pm 5.90$}&  $1.06$  {$\pm 0.34$} & 60 & 6.98  {$ \pm 1.45$}&  $1.11$  {$\pm 0.15$}& 40 & 42.59\\
        Ours & \cmark   & 3.55  {$\pm 0.70$}&  \textbf{0.70}  {$\mathbf{\pm 0.13}$} & $\mathbf{100}$ & \textbf{4.34  {$\mathbf{\pm 0.56}$}}& $\mathbf{0.89}$  {$\pm 0.11$} & $\mathbf{100}$ &{8.78}  \\
        % Ours & \cmark &  4.00  {$\pm 0.93$}&  $0.72$  {$\pm 0.11$} & $\mathbf{100}$ &4.97  {$\pm 1.23$}& \textbf{0.81}  {$\mathbf{\pm 0.15}$} & $\mathbf{100}$ & 36.59\\
\bottomrule
\end{tabular}
\label{tab:ablation}
\end{table*}

{\color{black} We compare the proposed method against two baselines for curvature estimation from the centerline point cloud. Both compute pointwise curvature values at the 
sample locations using finite difference approximations of derivatives and equation~\eqref{eq:curvature_form}. 
The \emph{naive} baseline applies this computation directly 
to the ordered 3D points. The \emph{smooth naive} baseline 
first applies a Savitzky-Golay filter~\cite{savitzky1964smoothing} 
with window length~5 before computing curvature via the 
same finite difference scheme. In both cases, the output 
is a discrete set of curvature values, not a continuous 
function of arc length. Obtaining a spatial curvature 
profile suitable for the MPC formulation would require 
interpolation, which does not directly provide bounded curvature (see Figure~\ref{fig:Smooth_est}), 
making these baselines unsuitable for direct integration 
with the  MPC formulation. Note that smoothing the curvature directly, rather than the centerline points, leads to poor and unpredictable Cartesian reconstruction error and is therefore not pursued. We evaluate the proposed method with and without initialization. In the latter case, only the number of sigmoids is retained from the initialization procedure, while their parameters are set to zero.

Table~\ref{tab:ablation} summarizes the results. The proposed method with initialization achieves the best overall performance: it matches the naive baseline in Cartesian reconstruction accuracy while substantially improving curvature estimation (lower MAE), and is the only configuration that achieves a 100\% success rate on both tracks. These results also confirm the critical role of initialization for the nonlinear optimization. Figure~\ref{fig:NORA_est} illustrates representative centerline estimates at different vehicle positions, showing that the fitted curvature model closely matches the ground-truth while remaining differentiable, ensuring bounded curvature, and hence compatibility with the Frenet-frame formulation.}

\subsubsection{\textcolor{black}{Real-World Validation}}
\label{subsec:curv_est_real}
{\color{black}

To validate the robustness of curvature estimation beyond the simulated setting, we evaluate it on real-world images collected from an F1TENTH vehicle equipped with an onboard Intel RealSense D405 RGB-D camera with $1280 \times 720$ resolution, at 30 frames per second. The vehicle was driven around a physical track in a laboratory environment, and RGB-D frames were recorded along with the vehicle pose obtained using a motion capture system. The track and an ego-centric view from the onboard camera are shown in Figure~\ref{fig:phys_track}. The track width is \SI{1}{m}, the centerline length is approximately \SI{20}{m}, and the curvature profile is piecewise-constant with varying levels along the track. While the real track shares the piecewise-constant curvature structure of the simulated ones, it is larger in scale and features a wider range of distinct curvature magnitudes.}

{\color{black} The curvature estimation pipeline was applied to the recorded
images offline, using the same algorithmic configuration as in
simulation, with several adjustments for the larger track scale
and noisier measurements: overlap tolerances were relaxed (proximity threshold increased from \SI{5}{cm} to \SI{8}{cm}, minimum overlap/novel point threshold reduced from 30 to 20), and the maximum map length was increased from \SI{3.5}{m} to \SI{3.8}{m}. All remaining parameters were kept identical to the simulated setting. This limited number of adjustments highlights the transferability of the proposed method across different track scales and sensing conditions.}

\textcolor{black}{ The raw centerline measurements exhibit a mean and worst Hausdorff distance from the ground-truth of \SI{15}{cm} and  \SI{1.5}{m}, respectively, driven by outliers in the edge detection and depth unprojection pipeline. These errors stem from multiple compounding 
sources: depth noise from the RGB-D sensor, imperfect lane markings 
(the tape has a width of \SI{4}{cm} and was applied manually), and, 
most critically, violation of the fixed-extrinsic calibration 
assumption due to suspension dynamics and mechanical vibrations.
% These
The optimization-based curvature estimation module substantially attenuates these disturbances. The mean Hausdorff distance between the estimated and ground-truth centerline is reduced to \SI{14}{cm}, the mean absolute curvature error is \SI{0.4}{m^{-1}}, and the worst-case Hausdorff distance is reduced to \SI{29}{cm}. Furthermore, the worst-case Hausdorff distance computed over only the first \SI{2}{m} of the map ahead of the vehicle, which is the most relevant segment for the receding-horizon controller, is \SI{22}{cm}, indicating that the estimation quality is highest precisely where it matters most for closed-loop performance.  These values are comparable to the simulated setting (Table~\ref{tab:ablation}) in comparison to the track width, confirming that the proposed method generalizes to real-world sensing conditions. 
Nevertheless, the residual estimation error cannot be neglected from a control perspective. The track half-width is \SI{50}{cm}, and the worst-case Hausdorff distance of \SI{29}{cm} constitutes approximately \SI{58}{\percent} of this margin. A nominal controller that treats the estimated centerline as ground-truth would therefore operate with a significantly reduced lateral safety budget, directly motivating the uncertainty-aware controller introduced in Section~\ref{sec:control_res}: since the residual geometric error cannot be further reduced at the perception level, it must be explicitly acknowledged and accounted for in the control formulation to maintain safety with respect to the lateral track constraints.}

\subsection{Sampling of Centerline Realizations}
\label{subsec:sampling_res}

For the estimates shown in Figure~\ref{fig:NORA_est}, we perform the sampling procedure described in Algorithm~\ref{al:2} and show the results in Figure~\ref{fig:sampling}. \textcolor{black}{The procedure is governed by three hyperparameters: the number of sampling attempts $N_{\text{rep}}$, the perturbation covariance $\Sigma_{\mathbf{\Theta}^k}$, and the Hausdorff distance bound $D^k$.} {\color{black} We set $D^k = \SI{10}{cm}$, to exceed the worst-case Hausdorff distance observed across all trajectories and tracks in the ablation study (Table~\ref{tab:ablation}). This provides a principled, data-driven upper bound on the perception-induced error. By varying $D^k$, one can directly control the conservatism of the uncertainty representation: smaller values yield a tighter set of admissible centerlines, while larger values account for less reliable perception. The covariance $\Sigma_{\mathbf{\Theta}^k} = 0.2$ was chosen to produce a meaningful spread of plausible parameter realizations while maintaining a sufficient acceptance rate among the sampled candidates. For visualization purposes, we set $N_{\text{rep}} = 500$ to clearly illustrate the variability induced by the uncertainty, and show all accepted realizations from $CandidateList$ in Figure~\ref{fig:sampling}. For the subsequent control evaluation in Section~\ref{sec:control_res}, this value is reduced to $N_{\text{rep}} = 100$, which was empirically found to still yield a sufficiently diverse set of samples at lower computational cost.}

\begin{figure}[t]
    \centering
    \includegraphics[width=0.99\linewidth]{ 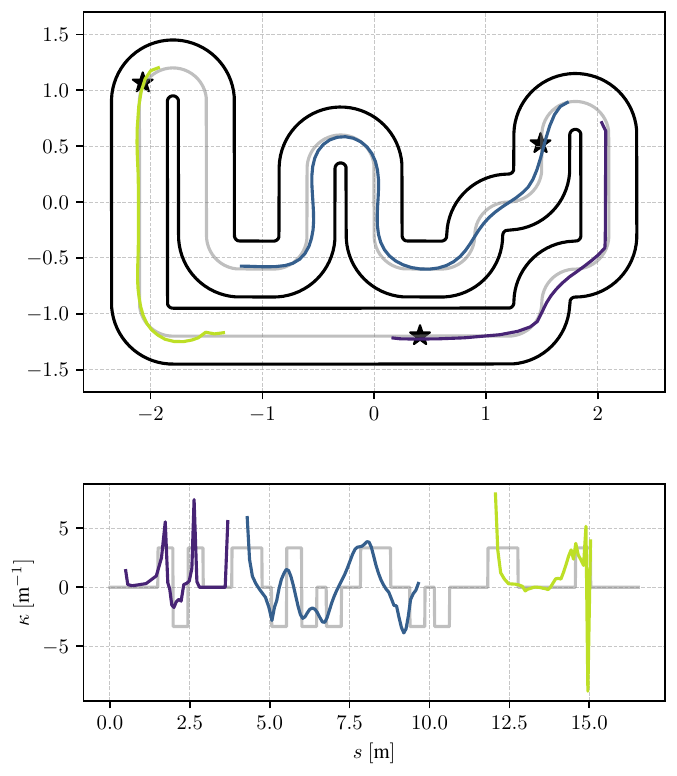}
    \caption{Top: results of the smooth naive method for centerline estimation on track B for three car positions (indicated by stars). Bottom: curvature estimates for each corresponding position on the track.}
    \label{fig:Smooth_est}
\end{figure}

\begin{figure}
    \centering
    \includegraphics[width=0.99\linewidth]{ 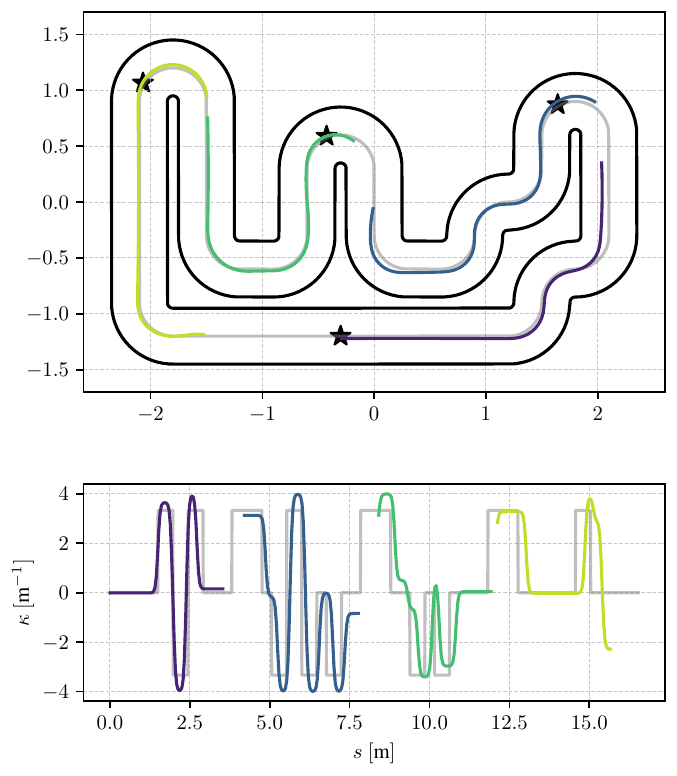}
    \caption{Top: results of the proposed method for centerline estimation on track B for three car positions (indicated by stars). Bottom: curvature estimates for each corresponding position on the track.}
    \label{fig:NORA_est}
\end{figure}

\begin{figure}
    \centering
    \includegraphics[width=0.99\linewidth]{ 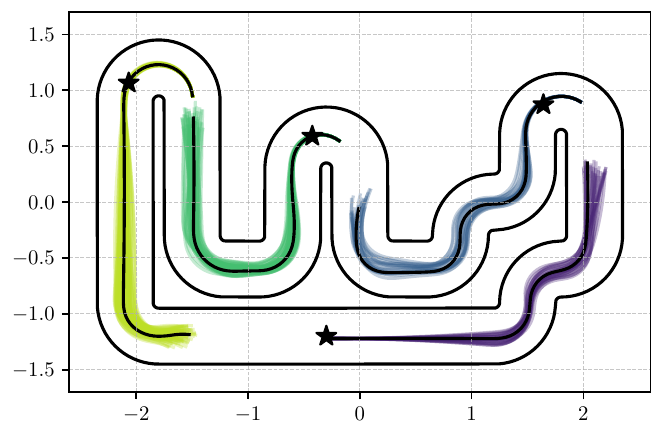}
    \caption{The results of sampling 500 centerline realizations according to Algorithm \ref{al:2}.}
    \label{fig:sampling}
\end{figure}

\subsection{\textcolor{black}{Uncertainty-Aware Control} }
\label{sec:control_res}

We incorporate the uncertainty-aware controller alongside the sampling procedure to solve the receding-horizon control problem defined in \eqref{eq:pa-control}. We evaluate three configurations: $N_s = 1$ (using only the estimated centerline), $N_s = 5$, and $N_s = 10$. \textcolor{black}{For $N_s > 1$, one realization corresponds to the 
estimated centerline, while the remaining $N_s-1$ are selected from the valid candidate set as those with the highest Hausdorff distance from the estimated centerline,  prioritizing centerlines 
that exhibit the largest pointwise spatial deviation from 
the nominal estimate.}

We choose the cost weights of the controller as follows: $q_s = 400$, $q_{\eta} = 100$, $q_{\phi} = 300$, $q_{v_x} = q_{v_y} = 5$, and $R = \text{diag}(0.01, 0.001)$ and all other parameters remain consistent with those used in the centerline estimation experiments.

{\color{black}Figure~\ref{fig:control} compares trajectories from the same initial condition under a nominal controller ($N_s = 1$) and the uncertainty-aware controller ($N_s = 10$). The uncertainty-aware controller produces more conservative behavior, steering the vehicle closer to the estimated centerline and away from regions where perception errors could lead to constraint violations. Notably, although the estimated centerline is geometrically close to the ground-truth (cf. Fig.~\ref{fig:NORA_est}), even small estimation errors can result in unsafe behavior when treated as exact. In the nominal case, the controller exploits the full lateral margin based on the estimated boundaries, which may lead to trajectories that approach or violate the true road limits. By contrast, the uncertainty-aware controller explicitly accounts for this mismatch and maintains a safety margin. Table~\ref{tab:ablation_control}(a) presents the results for varying numbers of centerline realizations. As $N_s$ increases, both the average and maximum lateral deviations from the ground-truth centerline are reduced, highlighting the benefit of explicitly accounting for perception uncertainty in the control formulation.}

{\color{black}To further characterize the role of $D^k$, we fix $N_s = 5$ 
as a favorable trade-off between practical robustness and computational 
cost, and vary $D^k \in 
\{5, 10, 15, 20\}\,$cm on Track~A. The upper end of the considered range corresponds 
to $80\%$ of the track half-width, beyond which the admissible 
set $\mathcal{M}^k$ spans nearly the entire road width and 
ceases to be informative. The results 
(Table~\ref{tab:ablation_control}(b)) confirm that the calibrated 
value $D^k = \SI{10}{cm}$ yields the lowest worst-case lateral 
deviation. Setting $D^k = \SI{5}{cm}$, which lies below the 
worst-case Hausdorff distance in Table~\ref{tab:ablation}, 
produces  maximum lateral deviation $\eta = \SI{22.56}{cm}$, approaching the track 
half-width comparably to the nominal controller  ($N_s = 1$). Increasing $D^k$ 
beyond $\SI{10}{cm}$ does not improve worst-case tracking, 
as maximum lateral deviation $ \eta$ rises back to $\SI{21.63}{cm}$ and 
$\SI{22.36}{cm}$ for $D^k = \SI{15}{cm}$ and $\SI{20}{cm}$, 
respectively. This suggests that an overly large admissible set 
introduces scenarios far from the nominal estimate, which can 
degrade the MPC solution at geometrically challenging track 
segments. Compared to the nominal controller, which 
achieves an average velocity of $\SI{1.49 }{m/s}$ on 
Track~A, the uncertainty-aware controller at the calibrated value $D^k = \SI{10}{cm}$
 operates at 
$\SI{1.47}{m/s}$, incurring less than $2\%$ velocity 
reduction while substantially improving worst-case lateral safety. 
These results highlight the data-driven calibration of $D^k$ and 
confirm that the formulation exhibits predictable sensitivity to 
this parameter.}

\begin{table}
    \caption{\textcolor{black}{\textbf{\textsc{Uncertainty-aware control.}}
    For each configuration, we run the uncertainty-aware controller 
    for 10 trajectories starting from different initial positions 
    and report the mean and maximal lateral distance from the 
    ground-truth centerline per lap, averaged over all trajectories 
    (avg $\eta$ and avg max $\eta$), the highest lateral distance 
    at all times (max $\eta$), and the mean Hausdorff distance 
    between the estimated and ground-truth centerline (avg HD).
    (a)~Sensitivity to the number of scenarios $N_s$ on Track~B 
    ($D^k = 10\,$cm).
    (b)~Sensitivity to the Hausdorff bound $D^k$ on Track~A 
    ($N_s = 5$).}}
    \label{tab:ablation_control}
    \setlength{\tabcolsep}{3pt}
    \centering
    %--- Panel (a) ---
    \smallskip
    {\small (a) Track B: varying $N_s$ \quad ($D^k = 10\,$cm)}\\[2pt]
    \begin{tabular}{|l|c|c|c|c|}
        \toprule
        {$N_s$} & avg $\eta$ [$cm$] $\downarrow$ & avg max $\eta$ [$cm$] $\downarrow$ & max $\eta$ [$cm$] $\downarrow$ & avg HD [$cm$] $\downarrow$ \\
        \midrule
        1  & ${4.37 \pm 0.85}$ & ${17.71 \pm 3.52}$ & ${22.61}$ & ${4.63 \pm 0.36}$ \\
        5  & ${3.48 \pm 0.29}$ & ${16.02 \pm 2.42}$ & ${21.08}$ & $\mathbf{4.47 \pm 0.47}$ \\
        10 & $\mathbf{3.41 \pm 0.34}$ & $\mathbf{14.07 \pm 2.25}$ & $\mathbf{17.57}$ & ${4.75 \pm 0.70}$ \\
        \bottomrule
    \end{tabular}

    \vspace{6pt}
    %--- Panel (b) ---
    {\color{black}{\small (b) Track A: varying $D^k$ \quad ($N_s = 5$)}\\[2pt]
    \resizebox{\columnwidth}{!}{\begin{tabular}{|l|c|c|c|c|}
        \toprule
        {$D^k$ [$cm$]} & avg $\eta$ [$cm$] $\downarrow$ & avg max $\eta$ [$cm$] $\downarrow$ & max $\eta$ [$cm$] $\downarrow$ & avg $v$ [$m/s$] $\uparrow$ \\
        \midrule
        5  & ${6.17 \pm 0.59}$ & ${17.91 \pm 2.16}$ & ${22.56}$ & ${1.48 \pm 0.02}$ \\
        10 & ${5.59 \pm 0.81}$ & ${16.95 \pm 1.35}$ & $\mathbf{18.95}$ & $\mathbf{1.47 \pm 0.02}$ \\
        15 & ${5.76 \pm 0.79}$ & ${17.21 \pm 2.16}$ & ${21.63}$ & ${1.45 \pm 0.01}$ \\
        20 & $\mathbf{5.22 \pm 1.12}$ & $\mathbf{15.53 \pm 3.03}$ & ${22.36}$ & ${1.45 \pm 0.07}$ \\
        \bottomrule
    \end{tabular}
    }}
\end{table}

{\color{black}

\begin{figure}[t]
    \centering
    \includegraphics[width=0.99\linewidth]{ 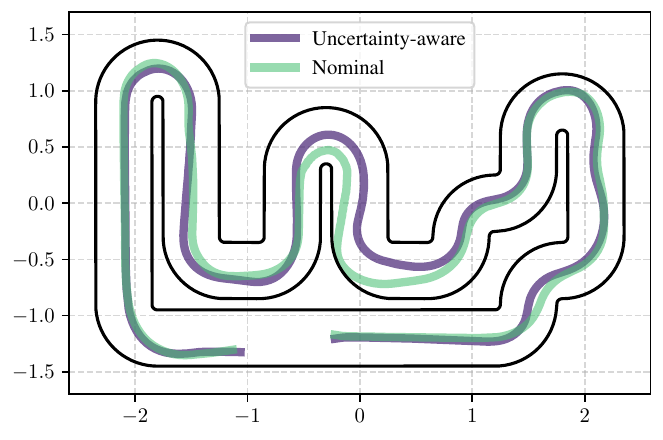}
    \caption{Comparison of trajectories of the nominal ($N_s=1$) and uncertainty-aware controller ($N_s=10$) that both rely on estimated centerline information.}
    \label{fig:control}
    \vspace{-0.5cm}
\end{figure}
\section{\textcolor{black}{Discussion and Limitations}}
\label{sec:Disc}

\textcolor{black}{This paper introduced an uncertainty-aware perception-based control framework that explicitly incorporates perception uncertainty into the decision-making process. Through simulation and real-world validation, we demonstrated that curvature-based centerline estimation enables accurate centerline reconstruction compatible with downstream control and that incorporating multiple centerline realizations into the control formulation integrates awareness of perception errors.}

\textbf{Guarantees and limitations:} The proposed controller does not provide formal worst-case or probabilistic safety guarantees. The term \emph{uncertainty-aware} indicates that geometric uncertainty is explicitly incorporated into the optimization, yielding empirical robustness with respect to bounded deviations in the reconstructed road model. The assumptions of a flat road and known width are not inherent to the
control formulation. If the perception module is extended to detect road
boundaries, the estimated width can be incorporated as a
progress-dependent lateral constraint, with estimation errors addressed
conservatively by constraining lateral deviation to the minimum reliably
estimated width along the prediction horizon. Similarly, elevation
changes can be accommodated by augmenting the geometric representation to
three dimensions. Only track banking would require a structurally
different road and vehicle model.

\textbf{Computational complexity:} The curvature estimation step is triggered only when 
sufficient new information is available, allowing 
it to run at a slower rate than the controller. Treating the sigmoid parameters as augmented states with identity dynamics, as is standard in moving horizon estimation with joint state and parameter estimation~\cite{kuhl2011real}, yields a multiple-shooting nonlinear program whose structure can be exploited by modern embedded MPC frameworks, e.g., {acados}~\cite{verschueren2022acados}, to solve problems comparable in size in real time. The scenario-based MPC exploits 
the Frenet-frame structure: the augmented state dimension per 
stage is $3N_s + 5$ rather than $8N_s$. For moderate values of $N_s$, the resulting 
problem dimensions remain within the range of nonlinear 
programs routinely solved in real time by~\cite{verschueren2022acados}. To validate 
real-time feasibility, we implemented the scenario MPC 
in \textcolor{black}{acados}~\cite{verschueren2022acados} and measured computation times on the 
F1TENTH platform (LattePanda Sigma, Intel Core i5-1340P). 
For $N_s \in \{1,5,10\}$, the average and maximum solve times using sequential quadratic programming with real-time iteration \textcolor{black}{over 6 laps on Track B} are $(\SI{4.6}{ms},\SI{13.2}{ms})$, $(\SI{8.1}{ms},\SI{15.0}{ms})$, and $(\SI{15.8}{ms},\SI{24.2}{ms})$, respectively, all below the \SI{33}{ms} sampling period. The 
simulation experiments in this paper use a Python/IPOPT 
prototype not optimized for real-time execution.

\textbf{Future work:} The known-pose assumption could be relaxed by estimating the vehicle state from 
onboard measurements. The lateral position and heading relative to the centerline 
can be recovered from a single RGB-D frame via standard geometric cues. The 
arc length coordinate could be obtained through point cloud registration (e.g., 
via the Iterative Closest Point algorithm) against the retained map segment, or by 
estimating relative arc length increments and jointly optimizing over them in~\eqref{eq:map_update}. 
Given the estimated spatial coordinates and the known vehicle dynamics, an Extended 
Kalman Filter could then be employed to recover the velocity states. The modular 
pipeline structure would allow such localization errors to be absorbed as additional 
geometric uncertainty within the existing Hausdorff-based framework, or treated as 
a separate bounded disturbance. More broadly, the proposed framework addresses 
perception-induced geometric uncertainty in the road 
reference, which is complementary to approaches that handle 
uncertainty in the vehicle dynamics and 
friction modeling~\cite{al2025lla, kalaria2024adaptive}. Since these 
two uncertainty sources act on different components of the 
system model, i.e., the reference path and the equations of 
motion, respectively, they could in principle be addressed 
jointly within a single control architecture. Finally, establishing 
formal robustness or probabilistic safety guarantees under perception-induced 
uncertainty remains an important open direction. 
}

\printbibliography 
\vspace{10\baselineskip}

\end{document}